\documentclass[10pt,twocolumn,journal]{IEEEtran}
\usepackage{graphicx}
\usepackage{booktabs}
\usepackage{bm}
\usepackage{dsfont}
\usepackage{mathrsfs}
\usepackage{authblk}
\usepackage{amsmath}
\usepackage{amssymb}
\usepackage{booktabs}
\usepackage{epstopdf}
\usepackage{subfigure}
\usepackage{caption}
\usepackage{textcomp}
\usepackage{color}
\usepackage{upgreek}
\usepackage{cite}
\usepackage{cleveref}
\usepackage[utf8]{inputenc}
\usepackage[english]{babel}
\usepackage{algorithm,algpseudocode}

\makeatletter
\def\thickhline{%
  \noalign{\ifnum0=`}\fi\hrule \@height \thickarrayrulewidth \futurelet
   \reserved@a\@xthickhline}
\def\@xthickhline{\ifx\reserved@a\thickhline
               \vskip\doublerulesep
               \vskip-\thickarrayrulewidth
             \fi
      \ifnum0=`{\fi}}
\makeatother

\newlength{\thickarrayrulewidth}
\DeclareMathOperator*{\argmax}{arg\,max}
\DeclareMathOperator*{\argmin}{arg\,min}

\begin{document}
\title{Multi-Antenna Coded Content Delivery with Caching: A Low-Complexity Solution}
\author{Junlin Zhao,~\IEEEmembership{Member,~IEEE,}
		Mohammad Mohammadi Amiri,~\IEEEmembership{Member,~IEEE,}
		and~Deniz G\"und\"uz, ~\IEEEmembership{Senior~Member,~IEEE}}
\maketitle

\begin{abstract}

We study downlink beamforming in a single-cell network with a multi-antenna base station serving cache-enabled users. Assuming a library of files with a common rate, we formulate the minimum transmit power with proactive caching and coded delivery as a non-convex optimization problem.
While this multiple multicast problem can be efficiently solved by successive convex approximation (SCA), the
complexity of the problem grows exponentially
with the number of subfiles delivered to each user in each time slot, which itself grows exponentially with the number of users. We introduce a low-complexity alternative through time-sharing that limits the number of subfiles received by a user in each  time  slot. We then consider the joint design of beamforming and content delivery with sparsity constraints to limit the number of subfiles received by a user in each time slot. Numerical simulations show that the low-complexity scheme has only a small  performance gap to that obtained by solving the joint problem with sparsity constraints, and outperforms state-of-the-art results at all signal-to-noise ratio (SNR) and rate values with a sufficient number of transmit antennas. A lower bound on the achievable degrees-of-freedom (DoF) of the low-complexity  scheme  is  derived  to characterize its performance in the high SNR regime.

\makeatletter{\renewcommand*{\@makefnmark}{}\footnotetext{Part of this work was presented at the IEEE International Workshop on Signal Processing Advances in Wireless Communications (SPAWC), Cannes, France, Jul. 2019 \cite{Zhao_spawc}.}\makeatother}
\makeatletter{\renewcommand*{\@makefnmark}{}\footnotetext{This work was partially supported by the European Research Council (ERC) through project BEACON (No. 677854), and by the European Union’s Horizon 2020 Research and Innovation Programme through project SCAVENGE (No. 675891).}\makeatother}
\makeatletter{\renewcommand*{\@makefnmark}{}\footnotetext{J. Zhao was with the Information Processing and Communications Lab, Department of Electrical and Electronic Engineering, Imperial College London, London SW7 2AZ, UK. He is now with the School of Science and Engineering, the Chinese University of Hong Kong, Shenzhen, Guangdong 518172, P.R. China, and also with the University of Science and Technology of China, Anhui 230026, P.R. China (e-mail: j.zhao15@imperial.ac.uk)}\makeatother}
\makeatletter{\renewcommand*{\@makefnmark}{}\footnotetext{M. Amiri was with the Information Processing and Communications Lab, Department of Electrical and Electronic Engineering, Imperial College London, London SW7 2AZ, UK. He is now with the Department of Electrical Engineering, Princeton University, Princeton, NJ 08544, USA (e-mail: mamiri@princeton.edu)}\makeatother}
\makeatletter{\renewcommand*{\@makefnmark}{}\footnotetext{D. G\"und\"uz is with the Information Processing and Communications Lab, Department of Electrical and Electronic Engineering, Imperial College London, London SW7 2AZ, UK (e-mail: d.gunduz@imperial.ac.uk)}\makeatother}

\end{abstract}
\IEEEpeerreviewmaketitle

\section{Introduction}

The seminal work of Maddah-Ali and Niesen showed that by exploiting caches at the users in order to create and exploit multicasting opportunities we can reduce the delivery time, or equivalently increase the throughput in wireless networks \cite{6763007}.
With coded caching, uncoded contents can be proactively pushed at user devices without knowing users' demands, and a server can serve multiple users simultaneously by broadcasting specially designed coded combinations of the remaining parts of all the users' requests, to guarantee that all the users can recover their desired contents.
This feature is particularly favorable
in wireless medium due to its broadcast nature.
However, \cite{6763007} ignored the physical characteristics of the channel, and simply assumed error-free communication, and focused on minimizing the number of bits delivered error-free over this multicast channel.

Over the last years many follow-up works have studied coded delivery over noisy broadcast channels.
When users may have different channel capacities, the user with the worst channel condition becomes the bottleneck limiting the performance of multicasting.
The global caching gain promised in \cite{6763007} is hence not straightforward in practice.
Coded caching in erasure broadcast channels is studied in \cite{8359316} and \cite{8036265} by allocating cache memories at weak receivers to overcome this bottleneck.
A simple binary Gaussian broadcast channel is considered in \cite{DBLP:journals/corr/ZhangE16b}, and an interference enhancement scheme is used to overcome the limitation of weak users.
A cache-aided multicasting strategy over a Gaussian broadcast channel is presented in \cite{MohammadDenizJSACPower}, with superposition coding and power allocation.
The authors in \cite{8254966} consider
fading channels, and show that a linear increase in the sum delivery rate with the number of users can be achieved with user selection.

Another important line of research has focused on evaluating the performance of coded caching and delivery in the presence of multiple transmit antennas at the server.
Multicast beamforming, where the multiple-antenna base station (BS) multicasts distinct data streams to multiple user groups, is an efficient physical layer technique \cite{Luo,4443878,Xiang}.
In \cite{Yang_Kobayashi_CC_fading}, the authors extend the results in \cite{8254966} to multi-input single-output (MISO) fading channels, where the same linear increase in content delivery rate with respect to (w.r.t.) the number of users is achieved without channel state information at the transmitter (CSIT), and an improvement is obtained with spatial multiplexing when CSIT is available.
In \cite{8581506}, coded delivery is employed along with zero-forcing to simultaneously exploit spatial multiplexing and caching gains. With multiple antennas at the BS, coded messages can be nulled at unintended user groups, which increases the number of users simultaneously served as compared to the single antenna setting.
Particularly, this approach was found to achieve the near-optimal degrees-of-freedom (DoF) in \cite{7580630}.
In addition to the gain in content delivery rate, employing multiple transmit antennas also allows reducing the
subpacketization level required in coded caching \cite{Elia_subpacket,salehi2019multi}.

By treating the transmission of coded subfiles as a coordinated beamforming problem, improved spectral efficiency is achieved in \cite{8950279} by optimizing the beamforming vectors,
which is also shown to achieve the same DoF as in \cite{8581506} in special cases.
Memory-sharing is proposed in  \cite{8007039} to apply the content placement scheme of \cite{6763007} for a fraction of the library, which exploits both the spatial multiplexing gain and the global caching gain by sending a common message together with user-dependent messages.
The impact of imperfect CSIT on achievable DoF is considered for MISO broadcast channels in \cite{Zhang_CSIT}.
Similarly to \cite{8007039}, \cite{8314745} adopts memory-sharing, and proposes a joint unicast and multicast beamforming approach.

In this paper, motivated by the results in \cite{8581506} and \cite{8950279}, we consider a cache-aided MISO broadcast channel.
Firstly, a general framework for cache-aided downlink beamforming is formulated, focusing on the minimum required transmit power for delivering the contents at a prescribed common rate. The resultant nonconvex optimization problem is tackled by successive convex approximation (SCA), which is guaranteed to converge to a stationary solution of the original nonconvex problem.
As noted in \cite{8950279}, the beamforming design involves solving an optimization problem with exponentially increasing number of constraints with the number of coded messages each user decodes in each time slot. To limit the complexity,
we propose a novel content delivery scheme, in which the coded subfiles, each targeted at a different subset of receivers, are delivered over multiple orthogonal time slots, while the number of coded messages each user decodes in each time slot can be flexibly adjusted.
Unlike the scheme in \cite{8950279}, the scheme we propose does not limit the number of users served in each time slot, but directly limits the number of messages each user decodes, and hence, the complexity of the decoder.
We propose a greedy algorithm that decides the multicast messages to be delivered at each time slot, and the number of time slots.
A lower bound on the DoF achieved by the greedy scheme is also provided.
We then consider a more general design of the beamforming vectors together with the content delivery scheme with a constraint on the maximum number of  messages each user can decode in any time slot. We formulate this joint optimization as a power minimization problem with sparsity constraints, and solve it via SCA to obtain a stationary solution.
Our numerical results show that the proposed greedy scheme has only a small performance gap to that of the optimization-based delivery scheme, and provide significant gains over the one proposed in \cite{8950279} in terms of transmit power, particularly in the high rate/high signal-to-noise ratio (SNR) regime.

The remainder of the paper is organized as follows. Section II introduces the system model. In Section III, we present an achievable coded delivery scheme, and formulate the power minimization problem for a multi-antenna server.
We introduce a low-complexity content delivery scheme in Section IV, and present its DoF analysis in Section V. In Section VI, we consider 
the joint design of beamforming and coded content delivery.
Finally, we compare the proposed schemes with the state-of-the-art through numerical simulations in Section VII, and conclude the paper in Section VIII.

\begin{figure}[!t]
\centering
\includegraphics[width=3.5in]{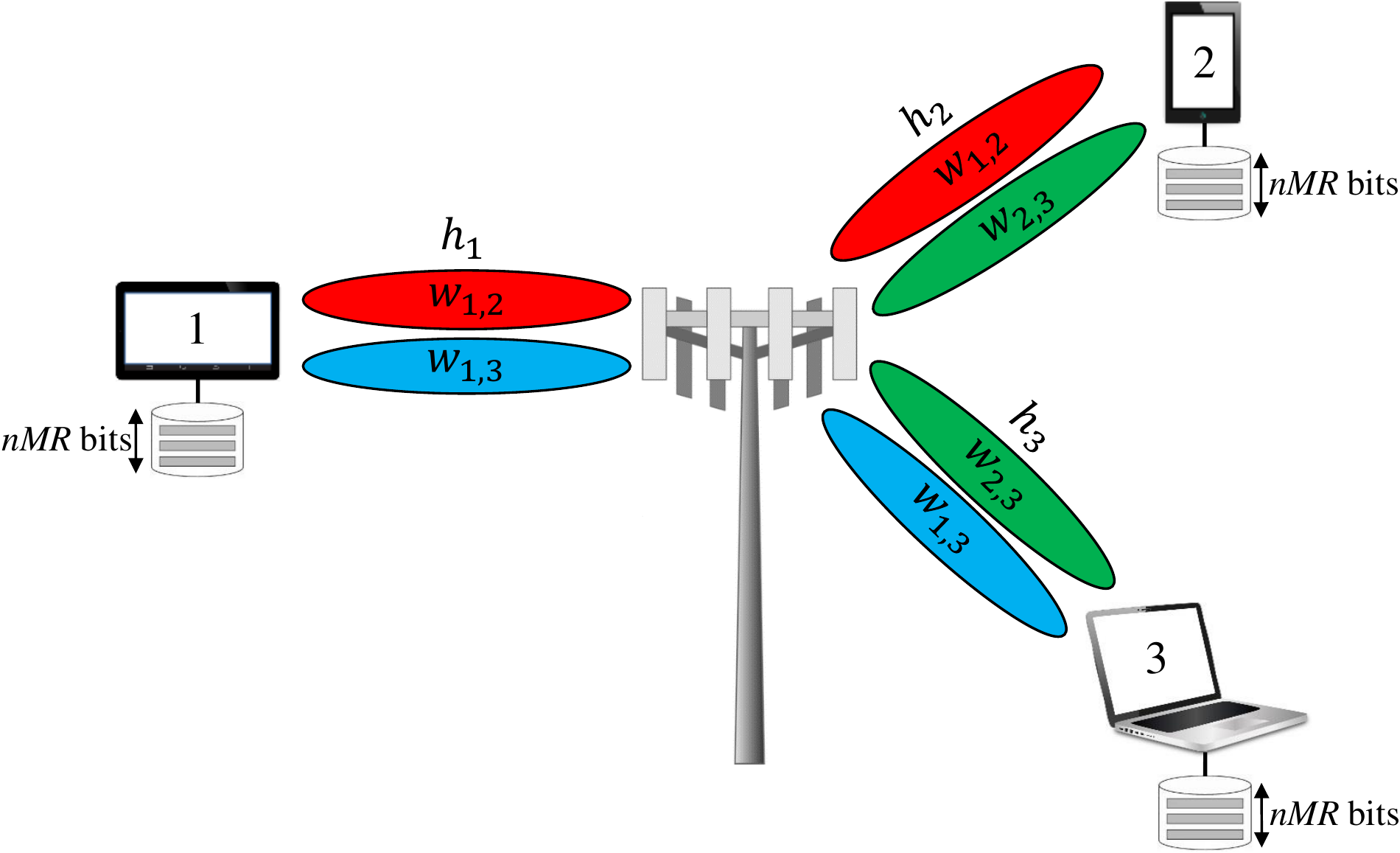}
\caption{Illustration of a cache-aided MISO channel with $K=3$ users. Multi-antenna BS employs multicast beamforming to deliver the missing parts of users' requests.}
\label{Sys_Mod}
\end{figure}


\section{System Model}\label{SystemModelSec}

We consider downlink transmission within a single cell, where a BS equipped with $N_T$ antennas serves $K$ single-antenna cache-equipped users, as illustrated in Fig.~\ref{Sys_Mod}. We consider a library of $N$ files, denoted by $ \boldsymbol{V} \triangleq (V_1,\cdots,V_N)$, each distributed uniformly over the set $
\left[ 2^{nR} \right]$\footnote{For any positive real $X$, we define $[X]$ as the set of positive integers less than or equal to X.}, available at the BS, where $R$ and $n$ represent the rate of each file and the blocklength, respectively. Each user is equipped with a local cache that can store up to $M$ files, and the corresponding \textit{caching factor}, $t$, is defined as the ratio of the total cache capacity across all the receivers to the library size, $t \triangleq MK/N$.

Contents are placed at users' caches during off-peak periods without any prior information on the user requests or the channel state information (CSI) to be experienced during the delivery phase. Caching function for user $k$ is denoted by $\phi_k^{(n)} : \left[ 2^{nR} \right]^N \to \left[ 2^{nMR} \right]$, which maps the library to the cache contents $Z_k$ at user $k$, i.e., $Z_k = \phi_k (\boldsymbol{V})$, $k \in [K]$. Once the users reveal their demands $\boldsymbol{d} \triangleq (d_1, \dots, d_K)$, where $d_k\in [N],\forall k\in [K]$, signal $\boldsymbol{X} \in \mathbb{C}^{N_T \times n}$ is transmitted, where $\boldsymbol{X} \triangleq [\bm{x}_1 \cdots \bm{x}_n]$, and $\bm{x}_i \in \mathbb{C}^{N_T \times 1}$ is the channel input vector at time $i,~i=1,\ldots,n$. An average power constraint $P$ is imposed on each channel input $\bm{X}$. User $k$ receives
\begin{align}\label{UserkSysModel}
\boldsymbol{y}_k^T = \boldsymbol{h}_k^H\boldsymbol{X}  + \boldsymbol{n}_k^{T},    
\end{align}
where $\bm{h}_k \in \mathbb{C}^{N_T \times 1}$ is the channel vector from the BS to the $k$-th user, and $\bm{n}_k \in \mathbb{C}^{{n \times 1}}$ is the additive white Gaussian noise at user $k$ with each entry independent and identically (i.i.d.) distributed according to $\mathcal{CN}(0,\sigma_k^2)$, $k \in [K]$. We assume that the CSI is perfectly known to the BS and the receivers in the delivery phase. Hence, the encoding function at the BS, $\psi^{(n)}: {\left[ 2^{nR} \right]^N} \times {\left[ N \right]^K} \times \mathbb{C}^{N_T \times K} \to \mathbb{C}^{N_T \times n}$, maps the library, the demand vector, and the CSI to the channel input vector. We note here that, while the channel encoding function $\psi^{(n)}$ depends on the demand vector and the CSI, caching functions $\phi_k^{(n)}$ depend only on the library. After receiving $\boldsymbol{y}_k$, user $k$ reconstructs $\hat{V}_{d_k}$ using its local cache content $Z_k$, channel vector $\bm{h}_k$, and
demand vector $\bm{d}$
through function $\mu_k^{(n)}: \mathbb{C}^n \times \left[ 2^{nMR} \right] \times \mathbb{C}^{N_T \times 1} \times \left[ N \right]^K \to \left[ 2^{nR} \right]$, i.e., $\hat{V}_{k} = \mu^{(n)}_k \left( \boldsymbol{y}_k, Z_k, \bm{h}_k, \bm{d} \right)$, $k \in [K]$. The probability of error is defined as ${\rm{P_e}} \triangleq \max_{\bm{d}} \max_{k \in [K]} {\rm{Pr}} \{ V_{d_k} \ne \hat{V}_k \}$.
An $(R,M,P)$ tuple is \textit{achievable} if there exist a sequence of caching functions ${\phi_1^{(n)}, \dots, \phi_K^{(n)}}$, encoding function $\psi^{(n)}$, and decoding functions ${\mu_1^{(n)}, \dots, \mu_K^{(n)}}$, such that ${\rm{P_e}} \to 0$ as $n \to \infty$. For file rate $R$ and cache size $M$, our goal is to characterize 
\begin{align}\label{AveragePowerMemoryTradeOff}
P^*\left( {R,M} \right) \buildrel \Delta \over = &\inf \left\{ { P:\left( {R,M, P} \right) \mbox{is achievable}} \right\},
\end{align}
which characterizes the minimum required transmit power that guarantees the reliable delivery of any demand vector.

\section{An Achievable Delivery Scheme}
In this section, we present a multi-antenna transmission scheme with coded caching, where the cache placement and coded content generation follows \cite{6763007},
while beamforming is employed at the BS to multicast coded subfiles to the receivers.

\subsection{Placement and Delivery Schemes}\label{SubSecPlaceDeliv}
For a caching factor $t \in \{ 1, \dots, K-1 \}$, we represent $t$-element subsets of $[K]$ by $\mathcal{G}^t_1, \dots, \mathcal{G}^t_{\binom{K}{t}}$. File $V_i$, $i \in [N]$, is divided equally into $\binom{K}{t}$ disjoint subfiles $V_{i,\mathcal{G}^t_1},\dots, V_{i,\mathcal{G}^t_{\binom{K}{{t}}}}$, each consisting of $n\frac{R}{\binom{K}{t}}$ bits. User $k$, $k \in [K]$, caches subfile $V_{i,\mathcal{G}^t_j}$, if $k \in \mathcal{G}^t_j$, $\forall j \in [ \binom{K}{t} ]$. The cache content of user $k$ is then given by $\bigcup\nolimits_{i \in [N]} {\bigcup\nolimits_{j \in [\binom{K}{t}]: k \in \mathcal{G}^t_j} {V_{i,\mathcal{G}^t_j}} }$.

During the \textit{delivery phase},
for any demand combination $\bm{d}$, we aim to deliver the coded message  
\begin{equation}\label{DefCodedDeliveredContentCentralized}
s_{\mathcal{G}_j^{{t}+1}} \buildrel \Delta \over = {{\bigoplus}_{k \in \mathcal{G}_j^{{t}+1}} {V_{{d_k},\mathcal{G}_j^{{t}+1}\backslash \{ k\} }}} 
\end{equation}
to all the users in set ${\mathcal{G}_j^{{t}+1}}$, for $j \in [ \binom{K}{t+1} ]$. Observe that, after receiving $s_{\mathcal{G}_j^{{t}+1}}$, each user $k \in \mathcal{G}_j^{{t}+1}$ can recover subfile $V_{{d_k},\mathcal{G}_j^{{t}+1}\backslash \{ k\} }$ having access to $V_{{d_l},\mathcal{G}_j^{{t}+1}\backslash \{ l\} }$, $\forall l \in \mathcal{G}_j^{{t}+1} \backslash \{k\}$.

We define $\mathcal{S} \triangleq \{\mathcal{G}^{t+1}_1, \dots, \mathcal{G}^{t+1}_{\binom{K}{t+1}}\}$ as the set of all the multicast messages, with each message $\mathcal{T}\in\mathcal{S}$ represented by the set of users it is targeting, and let $\mathcal{S}_k \subset \mathcal{S}$ denote the subset of messages targeting user $k$. We have $\left| \mathcal{S} \right| = \binom{K}{t+1}$ and $\left| \mathcal{S} _k\right| = \binom{K-1}{t}$.

The following examples will be used to better explain the proposed scheme:
\\
\indent\textbf{\textit{Example 1}}: Let $N=5,~K=5,~M=1$.\label{SubSecExample2}~We have $t=\frac{MK}{N}=1$. Each file is split into $\binom{K}{t}=5$ disjoint subfiles of the same size, where we represent file $i$, $i \in [N]$, as
\begin{align}
V_i = \left\{ V_{i, \{ 1 \}}, V_{i, \{ 2 \}}, V_{i, \{ 3 \}} V_{i, \{ 4 \}},V_{i, \{ 5 \}}\right\}.
\end{align}
The cache content of user $k$ is $Z_k = \cup_{i \in [N]} V_{i, \{k\}}$, $k \in [K]$,
which satisfies the cache capacity constraint. All user demands can be fulfilled by delivering the following $\binom{K}{t+1}=10$ subfiles:
{
\begin{align}
&s_{\{1,2\}} = V_{d_1,\{ 2 \}} \oplus V_{d_2,\{ 1 \}},~s_{\{1,3\}} = V_{d_1,\{ 3 \}} \oplus V_{d_3,\{ 1 \}},\nonumber\\
&s_{\{1,4\}} = V_{d_1,\{ 4 \}} \oplus V_{d_4,\{ 1 \}},~s_{\{1,5\}} = V_{d_1,\{ 5 \}} \oplus V_{d_5,\{ 1 \}},\nonumber\\
&s_{\{2,3\}} = V_{d_2,\{ 3 \}} \oplus V_{d_3,\{ 2 \}},~s_{\{2,4\}} = V_{d_2,\{ 4 \}} \oplus V_{d_4,\{ 2 \}},\nonumber\\
&s_{\{2,5\}} = V_{d_2,\{ 5 \}} \oplus V_{d_5,\{ 2 \}},~s_{\{3,4\}} = V_{d_3,\{ 4 \}} \oplus V_{d_4,\{ 3 \}},\nonumber\\
&s_{\{3,5\}} = V_{d_3,\{ 5 \}} \oplus V_{d_5,\{ 3 \}},~s_{\{4,5\}} = V_{d_4,\{ 5 \}} \oplus V_{d_5,\{ 4 \}}.\nonumber
\end{align}

}

\textbf{\textit{Example 2}}: Let $N=4,~K=4,~M=1$.\label{SubSecExample3}~We have $t=\frac{MK}{N}=1$. Each file is split into $\binom{K}{t}=4$ disjoint subfiles of the same size.
All user demands can be fulfilled by delivering the following $\binom{K}{t+1}=6$ subfiles:
{
\begin{align}
&s_{\{1,2\}} ,~s_{\{1,3\}},~
s_{\{1,4\}},~s_{\{2,3\}} ,~s_{\{2,4\}} ,~s_{\{3,4\}}.
\end{align}
Note that the message $s_\mathcal{T}$ is intended for users in set $\mathcal{T}$, but interferes with users in set $[K]\backslash \mathcal{T}$. Moreover, for any demand combination $\bm{d}$, all the users are required to decode the same number of messages, which is $\binom{K-1}{t}$.

}

\subsection{Multi-Antenna Transmission Scheme}

The delivery of the coded messages in set $\mathcal{S}$ to their respective receivers is a multi-antenna multi-message multicasting problem. Before introducing our low-complexity scheme in the next section, we present here a general transmission strategy based on message-splitting and time-division transmission. The messages in $\mathcal{S}$ can be transmitted over $B$ orthogonal time slots, the $i$-th of which is of
blocklength $n_i, i\in[B]$, where $\sum_{i=1}^Bn_i=n$.
The transmitted signal $\bm{X}(i) \triangleq [\bm{x}_{\sum_{j=1}^{i-1}n_j+1} \cdots \bm{x}_{\sum_{j=1}^{i}n_j}  ]$ at time slot $i \in [B]$ is given by
\begin{align}
    \bm{X}(i) = \sum\nolimits_{\mathcal{T} \in \mathcal{S}} \bm{w}_{\mathcal{T}}(i) \bm{s}_\mathcal{T}^T (i),
\end{align}
where $\bm{s}_\mathcal{T} (i)  \in \mathbb{C}^{n_i \times 1}$ is a unit-power complex Gaussian signal of blocklength $n_i$, modulated from the corresponding message $s_\mathcal{T}$ in (\ref{DefCodedDeliveredContentCentralized}), intended for the users in set $\mathcal{T}$, transmitted in time slot $i$, encoded by the beamforming vector $\bm{w}_\mathcal{T}(i)\in \mathbb{C}^{N_T \times 1}$.

The received signal at user $k$ in time slot $i$ is
\begin{align}\label{receivedSigUserk}
&\bm{y}_k^T(i) = \underbrace{\bm{h}_k^H \sum\limits_{\mathcal{T} \in \mathcal{S}_k} \bm{w}_{\mathcal{T}}(i) \bm{s}_\mathcal{T}^T (i)}_{\text{desired messages}} + \underbrace{\bm{h}_k^H \sum\limits_{\mathcal{I} \in \mathcal{S}_k^C} \bm{w}_{\mathcal{I}}(i) \bm{s}_\mathcal{I}^T (i)}_{\text{interference}} + \bm{n}_k^T(i),
\end{align}
where $\mathcal{S}_k^C$ is the complement of set $\mathcal{S}_k$ in $\mathcal{S}$. Let $\Pi_{\mathcal{S}_k}$ denote the collection of all non-empty subsets of $\mathcal{S}_k$, with each element of $\Pi_{\mathcal{S}_k}$ denoted by $\pi_{\mathcal{S}_k}^j, ~j\in[2^{\binom{K-1}{t}}-1]$. We denote $\mathcal{S}(i) \subset \mathcal{S}$ as the subset of messages transmitted in time slot $i$, i.e., $\mathcal{T}\in\mathcal{S}(i)$ if $\bm{w}_\mathcal{T}(i) \neq \bm{0}$.

Note that each user may receive more than one message in each transmission slot.
From the capacity region of the associated Gaussian multiple access channel, following conditions must be satisfied for successful decoding of all the intended messages at user $k$, $k \in [K]$, at time slot $i$:
\begin{align}\label{eq:capacity_constraint}
\sum\limits_{\mathcal{T} \in \pi_{\mathcal{S}_k}^j} R^\mathcal{T}(i) \leq \frac{n_i}{n}~\!\text{log}_2 \bigg(1+\sum\limits_{\mathcal{T} \in \pi_{\mathcal{S}_k}^j}\gamma_k^\mathcal{T}(i) \bigg),\;\forall \pi_{\mathcal{S}_k}^j \in \Pi_{\mathcal{S}_k},
\end{align}
where $R^\mathcal{T}(i)$ is the rate of message $\bm{s}_\mathcal{T}(i)$, and $\gamma_k^\mathcal{T}(i)$ is the received signal-to-interference-plus-noise ratio (SINR) of message $s_\mathcal{T}(i)$ at user $k$ at time slot $i$, given by
\begin{align}\label{eq:SINR}
\gamma_k^\mathcal{T}(i) \triangleq \frac{\vert \bm{h}_k^H \bm{w}_\mathcal{T}(i) \vert^2}{\sum\nolimits_{\mathcal{I} \in \mathcal{S}_k^C} \vert \bm{h}_k^H \bm{w}_\mathcal{I}(i) \vert^2+\sigma_k^2},
\end{align}
for any $ \mathcal{T} \ni k $, or equivalently, any $\mathcal{T}\in \mathcal{S}_k$.
The rate of message $\mathcal{T}$ is the sum of the rate of submessages $s_\mathcal{T}(i)$, and must satisfy
\begin{align}\label{eq:rate_constraint}
    \sum\nolimits_{i=1}^B R^\mathcal{T} (i) \geq \frac{R}{\binom{K}{t}},~\forall \mathcal{T}.
\end{align}

Note that this scheme is quite flexible; each multicast message can be split into $B$ messages and transmitted over $B$ time slots. It can be specialized to different content delivery schemes by specifying
the subset of transmitted subfiles in each time slot and the blocklength of each time slot, i.e., $\{\mathcal{S}(i)\}_{i=1}^B$ and $\{n_i\}$. Let
\begin{eqnarray}\label{eq:tdm_x}
v_\mathcal{T}(i) = 
\begin{cases}
1 & \text{if} \ \ \mathcal{T} \in \mathcal{S}(i) \\
0 & \text{if} \ \ \mathcal{T} \notin \mathcal{S}(i)
\end{cases}
\end{eqnarray}
be the indicator function specifying whether message $\mathcal{T}$ is transmitted at time slot $i$ or not. Note that $\Vert \bm{v}_\mathcal{T} \Vert_1 \geq 1$ is required to fulfill users' demands, where $ \bm{v}_\mathcal{T} \triangleq [ v_\mathcal{T}(1) \cdots v_\mathcal{T}(B) ]$.
It is readily seen that $v_\mathcal{T}(i)$ can be inferred by the corresponding beamforming vector $
\bm{w}_\mathcal{T}(i)$, or equivalently, by the message rate $R_\mathcal{T}(i)$.

\subsection{Transmit Power Minimization}
For any given delivery scheme specified by $v_\mathcal{T}(i)$ and $n_i$, $\forall i\in[B], \forall \mathcal{T}\in\mathcal{S}$,
the associated minimum required transmit power problem
is obtained as follows:
\begin{subequations}\label{eq:power}
\begin{flalign}
P \triangleq &\min\limits_{\{\bm{w}_\mathcal{T}(i)\},\{R^\mathcal{T}(i)\}} \ \ \!\!\!\! \sum\limits_{\mathcal{T} \in \mathcal{S}}\sum_{i=1}^B  \frac{n_i}{n} \Vert\bm{w}_{\mathcal{T}}(i)\Vert^2\\
&~~~~~~~\text{s.t.}~~~~~\sum\limits_{\mathcal{T} \in \pi_{\mathcal{S}_k}^j} R^\mathcal{T}(i) \leq \frac{n_i}{n}~\!\text{log}_2 \bigg(1+\sum\limits_{\mathcal{T} \in \pi_{\mathcal{S}_k}^j}\gamma_k^\mathcal{T}(i) \bigg),\nonumber\\
&~~~~~~~~~~~~~~~~~~~~~~~~~\;\forall \pi_{\mathcal{S}_k}^j \in \Pi_{\mathcal{S}_k},~i \in [B],~\forall k,\label{eq:constraint1}\\
&~~~~~~~~~~~~~~~~\sum\nolimits_{i=1}^B R^\mathcal{T} (i) \geq \frac{R}{\binom{K}{t}},~\forall \mathcal{T},\label{eq:constraint2}\\
&~~~~~~~~~~~~~~~~R^\mathcal{T}(i) \geq 0,~\forall i,\mathcal{T},\label{eq:scheme_constraint3}\\
&~~~~~~~~~~~~~~~~R^\mathcal{T}(i) = 0,~\text{if}~ v_\mathcal{T}(i) = 0,~\forall i,\label{eq:scheme_constraint}
\end{flalign}
\end{subequations}
where $\gamma_k^\mathcal{T}(i)$ is defined in (\ref{eq:SINR}). Here, constraints in (\ref{eq:constraint1}) guarantee that the rates of the messages targeting each user in each time slot are within the capacity region, constraints in (\ref{eq:constraint2}) 
ensure that sufficient information is delivered for each coded subfile over $B$ time slots, while (\ref{eq:scheme_constraint3})-(\ref{eq:scheme_constraint}) represent the specific content delivery scheme.

Note that the problem in (\ref{eq:power}) is a generalization of various well-known NP-hard problems depending on the specific content delivery scheme. For $B=|\mathcal{S}|$ with $|\mathcal{S}(i)|=1,~\forall i$, the problem boils down to a series of standard multicast beamforming problems, where a common message is broadcast to a different subset of $t+1$ users in each time slot \cite{Luo}. When $|\mathcal{S}(i)|>1$, $\mathcal{T}\bigcap \mathcal{T'}=\emptyset$ if $\mathcal{T}\neq \mathcal{T'}\in \mathcal{S}(i),~\forall i$, and $\mathcal{S}(i)\bigcap\mathcal{S}(j)=\emptyset$ if $i\neq j$, we need to solve the conventional multigroup multicast beamforming problem at each time slot \cite{4443878}.
It can be seen from (\ref{eq:power}) that the content delivery scheme specified by $v_\mathcal{T}(i)$ and $n_i$ affects the minimum required power.
A straightforward solution (as done in \cite{8007039},\cite{Vu_naive}) would transmit
a single coded message in each time slot. However, this does not exploit the spatial multiplexing gain provided by multiple antennas, and results in a poor DoF performance in the high SNR regime. Another approach studied in \cite{Vu_parallel} is to deliver the coded messages targeting non-overlapping user groups in parallel.
Obviously,
the content delivery scheme is an important factor on the system performance and needs to be carefully designed.

We remark here that, even when the delivery scheme is specified,
the problem in (\ref{eq:power}) is computationally intractable due to the non-convex constraints in (\ref{eq:constraint1}).
However, we show in the Appendix that SCA methods \cite{marks1978general} can be employed to obtain
a stationary point of the problem, which serves as
an upper bound on the optimal solution.
Starting with a feasible initial point, the SCA algorithm solves a sequence of subproblems in an iterative manner, where the subproblem in the $\nu$-th iteration is derived by convexifying the original problem at the solution point of the $(\nu-1)$-th subproblem. More detailed discussions on the SCA algorithm are provided in the Appendix.

\section{A Low-Complexity Design}\label{ProposedSchemeSec}

In this section, we propose a low-complexity content delivery scheme with the flexibility to adjust the number of coded messages intended for each user at each time slot.
Observing that if a set $\mathcal{S}(i) = \{ \mathcal{T}~\!\!|\!\!~v_\mathcal{T}(i) = 1 \}$ of messages are transmitted in time slot $i$,
$c_k(i) \triangleq | \mathcal{S}(i)  \bigcap  \mathcal{S}_k| $ messages are transmitted to user $k$, which results in $2^{c_k(i)}-1$ constraints only for user $k$ in time slot $i$ in problem (\ref{eq:power}).
Computational complexity of problem (\ref{eq:power}) increases drastically with the number of constraints, rendering the numerical optimization problem practically infeasible. More importantly, a multi-user detection scheme needs to be employed at the users, whose complexity also increases with $c_k(i)$.

A low complexity scheme is proposed in \cite{8950279} by limiting the number of users to be served in each time slot, thereby indirectly reducing the number of coded messages to be decoded by each user. Specifically, an integer parameter $\alpha \in [\text{min}\{N_T,K-t\}]$ is leveraged in \cite{8950279} to control the number of active users in each time slot, which is set to $t+\alpha$, and leads to a content delivery scheme with $B=\binom{K}{t+\alpha}$ time slots. In each time slot, a fraction of the desired coded messages for all the active users are transmitted.
In addition to $\alpha$, another integer parameter $\beta$ determines the possible set partitions of the user subset in each time slot. When $t+\alpha$ is divisible by $t+\beta$, the user subset can be partitioned into $\frac{t+\alpha}{t+\beta}$ non-overlapping subsets, and a fraction of the desired coded messages for each partition can be transmitted simultaneously.
It is shown in \cite{8950279} that the system performance can be improved if multiple groups of messages can be transmitted in parallel, i.e., $\frac{t+\alpha}{t+\beta}\geq 2$, as compared to the case $\beta=\alpha$.
Moreover, the number of messages for each user to decode in each time slot is $\binom{t+\beta-1}{t}$, which is an exponential function of $\beta$. Therefore, by adjusting the value of $\beta$, the number of coded messages for each user in each time slot is indirectly adjusted.

Instead of limiting the subsets of users to be served in each time slot, we propose to directly adjust the number of coded messages targeted to each user.
We will show that this results in a more efficient delivery scheme than the one in \cite{8950279}.
In Example~1, if we transmit all the messages in one time slot, i.e., $B=1$, a total of $|\mathcal{S}|=\binom{K}{t+1}=10$ coded subfiles are transmitted simultaneously, with each user decoding $\binom{K-1}{t}=4$ messages. Accordingly, in the optimization problem in (\ref{eq:power}) we will have $K \times (2^{|\mathcal{S}_k|}-1) = 75$ constraints.
To alleviate the computational complexity, the low complexity scheme in \cite{8950279} splits each subfile into $3$ minifiles, and the coded messages are grouped to serve a subset of $t+\alpha = 3$ users in each of the $B=\binom{K}{t+\alpha}=10$ time slots. Within each time slot, each user needs to decode 2 messages.
Note that the power minimization problem for each time slot can be solved independently; therefore, we would need to solve 10 smaller optimization problems, each with $3\times 3=9$ constraints.

In contrast, we propose to serve as many users as needed at each time slot while keeping $c_k(i)$ under a given threshold $s$ for each user $k$. In our Example 1, we can satisfy all the user requests in only $2$ time slots, by setting nonzero rate targets for the messages in
\begin{align}
\mathcal{S}(1) &= \{{\{1,2\},\{2,3\},\{3,4\},\{4,5\},\{1,5\}}\}, \mbox{ and}\nonumber\\
\mathcal{S}(2) &= \{{\{1,3\},\{2,4\},\{3,5\},\{1,4\},\{2,5\}}\}\nonumber
\end{align}
in time slots 1 and 2, respectively. Note that each user $k$ decodes only $c_k(i)=s= 2 $ messages in each time slot, the same as the delivery scheme in \cite{8950279}, requiring the same implementation complexity at each user; however, $5$ users are served in each time slot, which results in a significantly smaller number of time slots. Thus, we need to solve only two optimization problems at the BS, each with $5 \times 3=15$ constraints.

In general, the number of constraints in the optimization problem in \eqref{eq:power} increases exponentially with $s$, which results in exponentially increasing number of constraints in the problem in each SCA iteration. Thus the computational complexity of the delivery scheme can be largely alleviated by choosing a small $s$ value, which also simplifies the multi-user detection algorithm.

The key idea of our proposed low-complexity scheme is to divide set $\mathcal{S}$ into disjoint subsets $\mathcal{S}(1),\cdots,\mathcal{S}(B)$, with $c_k(i)\leq s$, $\forall k,i$, while keeping $B$ as small as possible. Since the total number of subfiles to transmit is fixed, choosing a small value of $B$, i.e., completing the delivery phase within a small number of time slots, requires multiplexing more messages in each time slot,
without increasing the complexity of the receivers.
To obtain this low-complexity scheme, the following optimization problem can be formulated:
\begin{subequations}\label{eq:opt2_delivery_original}
\begin{flalign}
&\min_{{\{\bm{v}_\mathcal{T}\},B}} \ \ B\\
&~~\text{s.t.} \  \sum\limits_{\mathcal{T} \ni k}v_\mathcal{T}(i) \leq s, \forall i\in[B],~k\in[K],\label{eq:opt2_delivery_c1}\\
&~~\ \ \ \ \sum\limits_{i=1}^B v_\mathcal{T}(i) = 1, \forall \mathcal{T},\label{eq:opt2_delivery_c2}\\
&~~\ \ \ \ v_\mathcal{T}(i) \in \{0,1\}, \forall \mathcal{T},i\in[B],
\end{flalign}
\end{subequations}
where constraint (\ref{eq:opt2_delivery_c1}) imposes that each user decodes no more than $s$ messages in each time slot, while (\ref{eq:opt2_delivery_c2}) requires that each message will be transmitted in only one time slot. However, since the problem itself varies with variable $B$,
the problem is not in a tractable form. By introducing $L\geq B$ as a prescribed parameter that determines the dimension of the problem, and an auxiliary variable $\bm{q}\in\{0,1\}^L$, problem (13) can be equivalently written as
\begin{subequations}\label{eq:opt2_delivery}
\begin{flalign}
B = &\min_{{\{\bm{v}_\mathcal{T}\},\bm{q}}} \ \ \bm{1}^T\bm{q}\\
&~~\text{s.t.} \  \sum_{\mathcal{T} \ni k}v_\mathcal{T}(i) \leq s, \forall i\in[L],~k\in[K],\\
&~~\ \ \ \ \sum_{i=1}^L v_\mathcal{T}(i) = 1, \forall \mathcal{T},\\
&~~\ \ \ \ \sum_{\mathcal{T}} v_\mathcal{T}(i) \leq \binom{K}{t+1} q_i, \forall i\in[L],\label{eq:opt2_delivery_c3}\\
&~~\ \ \ \ v_\mathcal{T}(i) \in \{0,1\}, \forall \mathcal{T},i\in[L],\label{eq:opt2_delivery_c4}\\
&~~\ \ \ \ \bm{q}\in \{0,1\}^L,
\end{flalign}
\end{subequations}
where $\bm{1}$ denotes a column vector of all ones.
Since $\binom{K}{t+1}$ is a bound on $\sum\nolimits_{\mathcal{T}} v_\mathcal{T}(i)$, the optimal $q_i$ is 1 if $\sum\nolimits_{\mathcal{T}} v_\mathcal{T}(i)$ is nonzero, and 0 otherwise, in order to minimize $\sum\limits_{i=1}^L q_i$ in the objective.
Note that problem (14) can be considered as minimizing the number of time slots employed out of a maximum $L$ available time slots. We can set $L = \binom{K}{t+1}$ which guarantees the existence of a solution; however, choosing a smaller $L$ will reduce the complexity of the problem.
The problem in (\ref{eq:opt2_delivery}) is a $0-1$ integer programming problem, which is generally NP-hard \cite{10.1007/978-3-642-95322-4_17}.

\begin{algorithm}[H]\label{alg:proposed}
\caption{Low-complexity greedy delivery scheme}
\label{alg:loop}
\begin{algorithmic}[1]
\Require{$N,K,M,s,R$} 
\Ensure{$B$,~$\bigcup_{i=1}^B \{S(i)\}$,~$\bigcup_{i=1}^B \{n_i\},~\forall \mathcal{T}$}
\State Set $t=\frac{MK}{N}$, $i=1$, and $\mathcal{E}=\mathcal{S}$

    \While {$\mathcal{E}\neq \varnothing$}
    	\State Set $\bm{c}(i)\triangleq [c_1(i)\cdots c_{K}(i)]=\bm{0}$, $\mathcal{S}(i)=\varnothing$, $\mathcal{C}=\mathcal{E}$
    	\While {$c_k(i) \leq s,\forall k\in[K]$ and $\mathcal{C}\neq \varnothing$}
    		\State $\mathcal{K}\triangleq \{ k~\!|\!~\argmin\limits_k \bm{c}(i)\}$
    		\State Find $\hat{\mathcal{T}}=\argmax\limits_{\mathcal{T}\in \mathcal{C}} |\mathcal{K}\bigcap\mathcal{T}|$
    		\State $\mathcal{C} = \mathcal{C} \backslash \{\mathcal{\hat{T}}\}$
    		\If{$c_k(i)+1 \leq s, \forall k\in\mathcal{\hat{T}}$}
    		    \State $c_k(i) = c_k(i)+1, \forall k\in\hat{\mathcal{T}}$
    			\State $\mathcal{S}(i) = \mathcal{S}(i) \bigcup \mathcal{\hat{T}}$, $\mathcal{E} = \mathcal{E} \backslash \{\mathcal{\hat{T}}\}$
    		\Else \State \textbf{break}
    		\EndIf
    	\EndWhile
    	\State $i\leftarrow i+1$
    \EndWhile
\State Set $B = i-1$
\For {$i=1:B$}
    \State $n_i=\frac{|\mathcal{S}(i)|}{\binom{K+1}{t}}n$
    \State  \begin{subequations}
            $R^\mathcal{T}\!(i)=\!
            \begin{cases}
            \frac{R}{\binom{K}{t}},
            ~\forall~\!\mathcal{T}\in\mathcal{S}(i)\\
            0,~\text{otherwise}
            \end{cases}$
            \end{subequations}
\EndFor
\end{algorithmic}
\label{alg:proposed}
\end{algorithm}

In Algorithm~\ref{alg:proposed}, we propose a greedy solution that constructs disjoint $\mathcal{S}(i)$ sets for any $s$ value. Specifically, $\mathcal{S}(i)$'s are generated in a sequential manner:
to construct $\mathcal{S}(i)$, we initialize $\bm{c} (i) \triangleq[c_1(i) \cdots c_K(i)] = \bm{0}$, $\mathcal{S}(i)=\varnothing$, and the set $\mathcal{E} = \mathcal{S}\backslash \bigcup_{j=1}^{i-1}\mathcal{S}(j)$ of remaining messages for assignment, we first identify the user(s) that have decoded the least number of messages so far, i.e., user(s) in set $\mathcal{K} \triangleq \{ k~\!\!|\!\!~\argmin_k \bm{c}(i)\}$, and check whether there exists a message $\mathcal{\hat{T}}\in\mathcal{E}$ such that the condition $c_k(i)+1 \leq s$ for $\forall k \in \mathcal{\hat{T}}$ holds.
If no such $\mathcal{\hat{T}}$ can be found, the process of constructing $\mathcal{S}(i)$ is completed, and we start constructing $\mathcal{S}(i+1)$ in the same manner.
The whole procedure is completed when $\mathcal{E} = \varnothing$, i.e., all the messages have been assigned to a subset.
Note that our proposed greedy scheme covers the case of $B=1$, where all the messages are sent simultaneously, if $s \geq \binom{K-1}{t}$.

Next we elaborate the proposed greedy content delivery algorithm in Examples~1 and 2.\\
\indent\textit{\textbf{Example 1 }(continued)}: $N=5,~K=5,~M=1$.\label{SubSecExample2}~$t=\frac{MK}{N}=1$. Suppose $s=2$.
\begin{figure*}[!tp]
\centering
\includegraphics[width=6.2in]{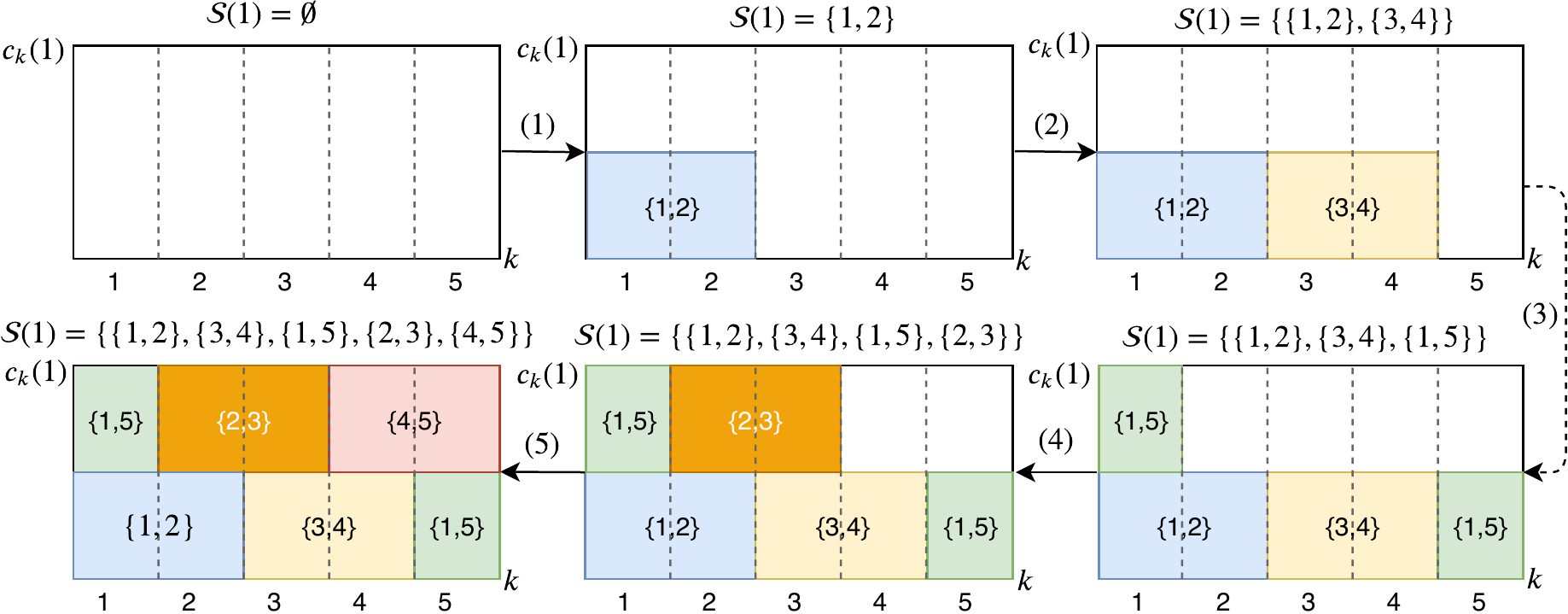}
\caption{Illustration of the proposed low-complexity greedy scheme for the network with $N=K=5$, and $M=1$.}
\label{Scheme_exp}
\end{figure*}
As illustrated in Fig.~\ref{Scheme_exp}, the algorithm starts by constructing $\mathcal{S}(1)$, i.e., identifying the coded messages to be delivered in the first time slot, initialized as $\bm{c} (1) \triangleq[c_1(1) \cdots c_K(1)] = \bm{0}$, $\mathcal{S}(1)=\varnothing$. Firstly, it is obvious that $\mathcal{K} = [K]$ since $c_k(1)=0,~\forall k$. Hence, one may choose any of the available messages in $\mathcal{E}=\mathcal{S}$. Suppose message $s_{\{1,2\}}$ is chosen. We update $c_1(1) = c_2(1) = 1$, $\mathcal{E} = \mathcal{E} \backslash \{1,2\}$. The algorithm then identifies the updated $\mathcal{K} = [3,4,5]$, according to which one may choose from $s_{\{3,4\}},s_{\{3,5\}},s_{\{4,5\}}$ without violating the constraint $c_k(1)+1\leq s$,  $\forall k$. Suppose $s_{\{3,4\}}$ is chosen, and we have $c_1(1) = c_2(1)= c_3(1) = c_4(1) = 1$,  $\mathcal{E} = \mathcal{E} \backslash \{3,4\}$, and $\mathcal{K}=[5]$; and accordingly one may choose from $s_{\{1,5\}},s_{\{2,5\}},s_{\{3,5\}},s_{\{4,5\}}$, while still keeping the constraint $c_k(1)+1\leq s$, $\forall k$. Similarly, messages $s_{\{2,3\}}$ and $s_{\{4,5\}}$ can be chosen, and the algorithm for $\mathcal{S}(1)$ is completed since $c_k(1)=2$, $\forall k$, and adding any of the remaining messages will violate the constraint. The algorithm then turns to construct $\mathcal{S}(2)$ similarly, until all the messages have been chosen, i.e., $\mathcal{E}=\emptyset$.

{\textit{Remark 1}}: As it can be seen above, the content delivery scheme obtained via Algorithm~\ref{alg:proposed} is not unique.
For instance, another feasible content delivery scheme with $s=2$ for Example 1 is:
\begin{align}
\mathcal{S}(1) &= \{{\{1,2\},\{3,4\},\{1,5\},\{2,4\},\{4,5\}}\}, \mbox{and}\nonumber\\
\mathcal{S}(2) &= \{{\{1,3\},\{2,3\},\{3,5\},\{1,4\},\{2,5\}}\}.\nonumber
\end{align}

\textbf{\textit{Example 2}}: $N=4,~K=4,~M=1$. $t=\frac{MN}{K}=1$.\label{SubSecExample2}

We present the content delivery schemes obtained from Algorithm~\ref{alg:proposed} for different values of $s$:\\
\textbf{Case 1: $\bm{s=1}$}. Algorithm~1 leads to $B=3$ time slots, each with $\frac{1}{3}n$ channel uses, and
\begin{align}
\mathcal{S}(1) &= \{{\{1,2\},\{3,4\}}\},\nonumber\\
\mathcal{S}(2) &= \{{\{1,3\},\{2,4\}}\},\nonumber\\
\mathcal{S}(3) &=
\{{\{1,4\},\{2,3\}}\}.\nonumber
\end{align}
It is noted that the scheme is the same as the one in \cite{8950279} obtained for $\alpha = 3~\text{and}~\beta=1$, in the sense that the same sets of messages are transmitted over the same number of time slots.\\
\textbf{Case 2: $\bm{s=2}$}. Algorithm 1 leads to $B=2$ time slots, and
\begin{align}
\mathcal{S}(1) &= \{{\{1,2\},\{3,4\}}\},\{{\{1,3\},\{2,4\}}\}, \nonumber\\
\mathcal{S}(2) &=
\{{\{1,4\},\{2,3\}}\}.\nonumber
\end{align}
It is noted that there is no $\beta$  value to induce a scheme in \cite{8950279} in this scenario, since $\beta$ can only take the value of $2$ to have $s=2$, making $t+\alpha$ not divisible by $t+\beta$.

{\textit{Remark 2}}: The proposed greedy content delivery scheme can be easily extended by limiting the number of active users as in \cite{8950279}. Specifically, instead of serving as many users as possible, which is up to $K$, Algorithm~\ref{alg:proposed} can be applied for a user subset of size $t+\alpha$ to obtain a content delivery scheme under the constraints on the number of messages to decode. While $\frac{t+\alpha}{t+\beta}$ must be an integer to induce a content delivery scheme in \cite{8950279}, Algorithm~\ref{alg:proposed} always provides a delivery scheme for any $s$ value. Therefore, our proposed greedy scheme can be considered as a generalization of the one in \cite{8950279}.

{\textit{Remark 3}}: It is noted that  Algorithm~\ref{alg:proposed} may lead to unequal number of messages transmitted in different time slots, which can be highly sub-optimal. An intuitive way to enhance the performance is to allocate more channel uses to the time slot with more messages to deliver.
In general, once the non-overlapping partition of $\mathcal{S}$, i.e., $\bigcup_{i=1}^B\{\mathcal{S}(i)\}$, is obtained, we can set the blocklength for the transmission of $\mathcal{S}(i)$ proportionally to the number of messages $|\mathcal{S}(i)|$.
For instance, in the case of $s=2$ in Example 2, we can allocate $\frac{2n}{3}$ channel uses for $\mathcal{S}(1)$ and $\frac{n}{3}$ channel uses for $\mathcal{S}(2)$.

\textit{Remark 4}: The proposed low-complexity scheme focuses on limiting $s$, which determines the complexity of multiuser detection at the receivers. On the other hand, the beamformer design resorts to solving the optimization problem in (12), whose computational complexity depends on the content delivery scheme.
Given the proposed low-complexity scheme in Algorithm~\ref{alg:proposed}, the optimization problem in (12) can be solved as detailed in the Appendix, which is firstly decomposed into at most $B_u$ problems in the form of (\ref{eq:optimization_i}).
For each problem, we have at most $K(2^s-1)$ constraints to
characterize the achievable rate region.
For the same $s$ value, the optimization problem for the scheme in \cite{8950279} can be decomposed into $B_l \triangleq \binom{K}{t+\alpha}\frac{(t+\alpha)!}{\delta!(t+\beta)!^\delta}$ subproblems, each with $(t+\alpha)(2^s-1)$ constraints for the achievable rate region characterization, where $\delta = \frac{t+\alpha}{t+\beta}$.
Therefore, each optimization problem in the proposed scheme involves $\frac{K}{t+\alpha}$ times more constraints compared to \cite{8950279}, but may require solving much fewer problems.

\begin{table*}[!ht]\caption{Comparison between Algorithm~\ref{alg:proposed} and \cite{8950279} in terms of the number of constraints they need to consider for each optimization problem, and the number of time slots, or equivalently, the number of optimization problems to be solved.}\label{tab:Num_constraints}
\centering
\begin{tabular}{ccccc|ccccc|ccccc}
\multicolumn{5}{c|}{$(K,t,s)=(10,1,1)$} & \multicolumn{5}{|c|}{$(K,t,s)=(10,2,1)$} &
\multicolumn{5}{|c}{$(K,t,s)=(10,3,1)$}\\
\thickhline
$(\beta,\alpha)$   & $\frac{K}{t+s}$   & $B_u$   & $B_l$   & $\frac{B_u}{B_l}$& $(\beta,\alpha)$  & $\frac{K}{t+s}$   & $B_u$   & $B_l$ & $\frac{B_u}{B_l}$  & $(\beta,\alpha)$   & $\frac{K}{t+s}$   & $B_u$   & $B_l$ & $\frac{B_u}{B_l}$\\
\hline
$(1,1)$ & $5$ & $9$ & $45$ & 0.2 &  $(1,1)$ & $\frac{10}{3}$ & $40$ & $120$ &$0.3333$ &$(1,1)$ & $\frac{5}{2}$ & $105$ & $210$ & 0.5\\
$(1,3)$ & $\frac{5}{2 }$ & $9$ & $630$ &$0.0143$ & $(1,4)$ & $2$ & $40$ & $2100$ &$0.0190$ &$(1,5)$ & $\frac{5}{4}$ & $105$ & $1575$ &$0.0667$\\
$(1,5)$ & $\frac{5}{3 }$ & $9$ & $3150$ &$0.0029$ & $(1,7)$ & $\frac{10}{7}$ & $40$ & $2800$ & $0.0143$\\
$(1,7)$ & $\frac{5}{4 }$ & $9$ & $4725$ &$0.0019$  &  & & \\
$(1,9)$ & $1$ & $9$ & $945$ &$0.0095$&  &  & & \\
\end{tabular}
\end{table*}

Table~\ref{tab:Num_constraints} compares the proposed scheme in Algorithm~\ref{alg:proposed} and the scheme in [16] in terms of the number of time slots, or equivalently, the number of optimization problems to be solved, denoted by $B_u$ and $B_l$, respectively, and the ratio of the number of constraints in each of these problems in the former scheme to the latter. We observe that, by simultaneously serving as many users as possible, the optimization problem in each time slot of the proposed low-complexity scheme has comparable number of constraints, but much fewer such optimization problems are needed, as compared to the scheme in \cite{8950279}.

Numerical results for the minimum required power for the proposed greedy transmission scheme, and the comparison with the one proposed in \cite{8950279} will be presented in Section VII.

\section{Degrees-of-Freedom (DoF) Analysis}\label{SecDoF}
In this section we analyze the performance of the scheme proposed in Algorithm \ref{alg:loop} in terms of the DoF it achieves in the high SNR regime.
To this end, we develop a content delivery scheme which upper bounds the number of time slots $B$ of the scheme presented in Algorithm \ref{alg:loop}.

For caching factor $t=MK/N$, Algorithm \ref{alg:loop} constructs, at each time slot, a set of distinct subsets of users of size $t+1$, such that no user index appears more than $s$ times in the set. This procedure is repeated until all the $\binom{K}{t+1}$ distinct subsets of users are selected exactly once.
While Algorithm \ref{alg:loop}
aims at minimizing the number of time slots required to deliver all the multicast messages, it is not possible to know in advance how many time slots will be needed.
To overcome this uncertainty, we
develop a more relaxed scheme, which utilizes a higher number of time slots than the one presented in Algorithm \ref{alg:loop}. 

We assume that the BS is equipped with $N_T \ge K-t$ antennas, to simultaneously transmit each coded packet of rate $R\slash \binom{K}{t}$ to the $t+1$ users that are interested in this message, while zero-forcing it at the remaining $K-t-1$ users \cite{7580630}.
At each time slot all the $K$ users are targeted, and each user receives no more than $s$ coded packets, and the coded delivery is performed for a total of $B$ time slots.
Assuming equal blocklength for different time slots in the high SNR regime, i.e., $n_i = n/B$, $\forall i \in [B]$, since $N_T \ge K-t$, we can lower bound the per-user DoF as follows \cite{IntAlignCadambeJafar,RealInterAlignAbol}:
\begin{align}\label{DoFLowerBoundAlg}
\mathrm{DoF} \ge \frac{\binom{K}{t}}{sB},    
\end{align}
where $B$ is the number of time slots obtained from Algorithm \ref{alg:loop}.

Next, we present an upper bound on $B$. We further divide each time slot to $s$ sub-time-slots for the new content delivery approach. At each sub-time-slot, the goal is to create a set of $(t+1)$-element subsets of users,
such that no user index appears in more than one subset. Due to the symmetry, it is easy to verify that the maximum number of such subsets at each sub-time-slot is $\left\lfloor {\frac{K}{{t + 1}}} \right\rfloor$. Therefore, we can generate a set of distinct $(t+1)$-element subsets of users, with no user appearing more than $s$ times by repeating this procedure for $s$ sub-time-slots. Accordingly, we can generate a set of $s\left\lfloor {\frac{K}{{t + 1}}} \right\rfloor$ distinct subsets of users, each of size $t+1$, such that each user index appears no more than $s$ times. This provides us with the set of coded packets for delivery at each time slot. 
For example, considering Example 1, where $N=5,~K=5,~t=1$, and $s=2$, we can generate the following sets of subsets of users, each of size $2$, in the two sub-time-slots of the first time slot:
\begin{align}
\mathcal{S}(1,1) &= \{\{1,2\},\{3,4\}\}, \mbox{and}\nonumber \\
\mathcal{S}(1,2) &= \{\{1,3\},\{4,5\}\},
\end{align}
where $\mathcal{S}(1) = \{ \mathcal{S}(1,1), \mathcal{S}(1,2) \}$, and this partitioning is not unique.
We exploit the symmetry in the subsets of users of size $t+1$, to which the coded packets are targeted, to obtain the total number of time slots required for transmission. By creating $s\left\lfloor {\frac{K}{{t + 1}}} \right\rfloor$ $(t+1)$-element distinct subsets of users at each time slot, we need no more than $\left\lceil \frac{\binom{K}{t+1}}{s\left\lfloor {\frac{K}{{t + 1}}} \right\rfloor} \right\rceil$
time slots to deliver all the coded messages. 

We highlight that, this approach resembles the set generation process in Algorithm \ref{alg:proposed}, but is stricter as it requires each user index to appear no more than once in each sub-time-slot, compared to Line 8 in Algorithm \ref{alg:proposed}. Hence, it results in fewer selected messages at each time slot, and more time slots.
\begin{figure*}[htbp]
\centering
\subfigure[$N=K=8,M=1.$]{
\begin{minipage}[t]{0.33\linewidth}
\centering
\includegraphics[scale=0.3]{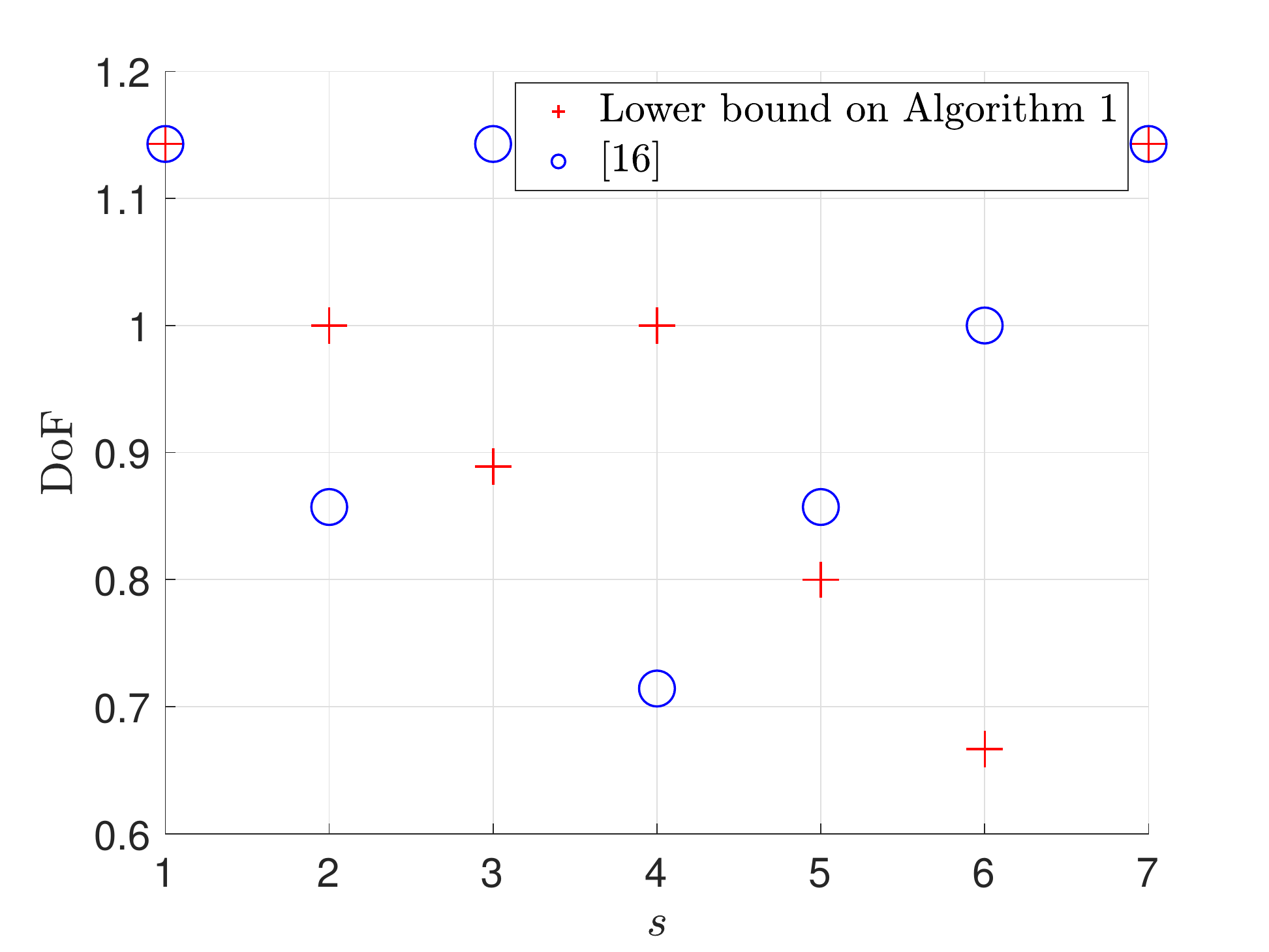}
\end{minipage}%
}%
\subfigure[$N=K=8,M=2.$]{
\begin{minipage}[t]{0.33\linewidth}
\centering
\includegraphics[scale=0.3]{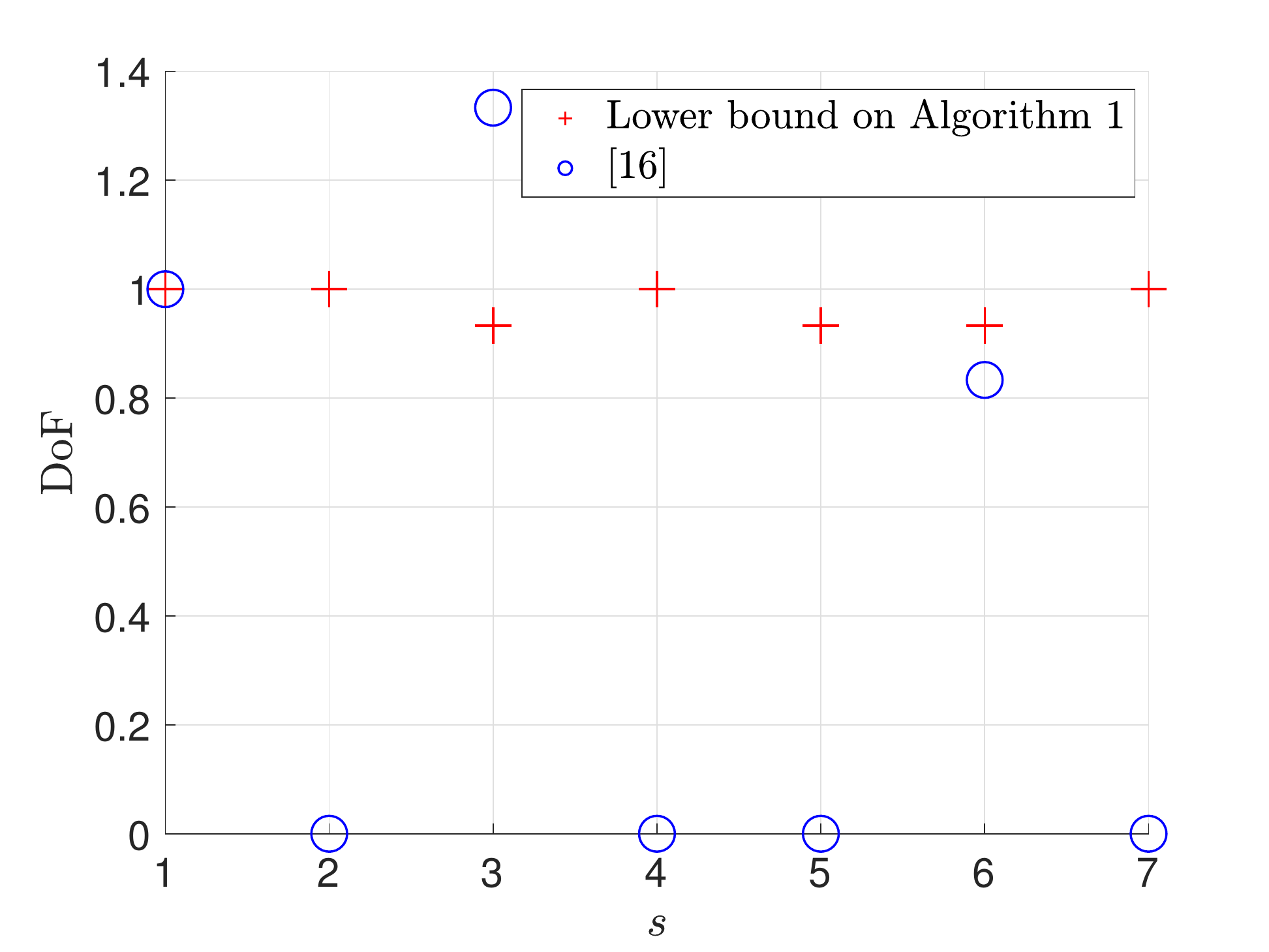}
\end{minipage}%
}%
\centering
\subfigure[$N=K=8,M=3.$]{
\begin{minipage}[t]{0.33\linewidth}
\centering
\includegraphics[scale=0.3]{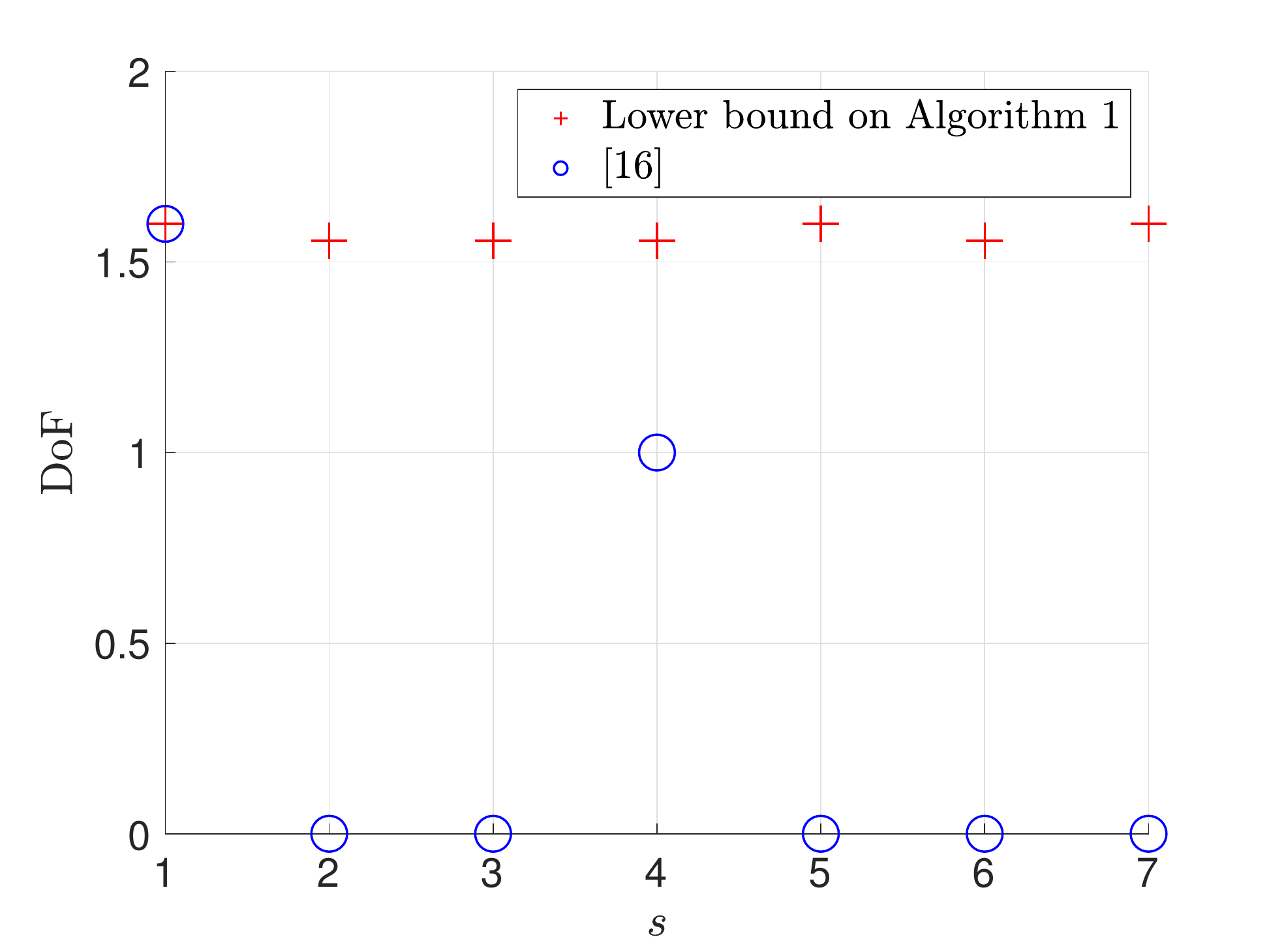}
\end{minipage}%
}%

\centering
\subfigure[$N=K=9,M=1.$]{
\begin{minipage}[t]{0.33\linewidth}
\centering
\includegraphics[scale=0.3]{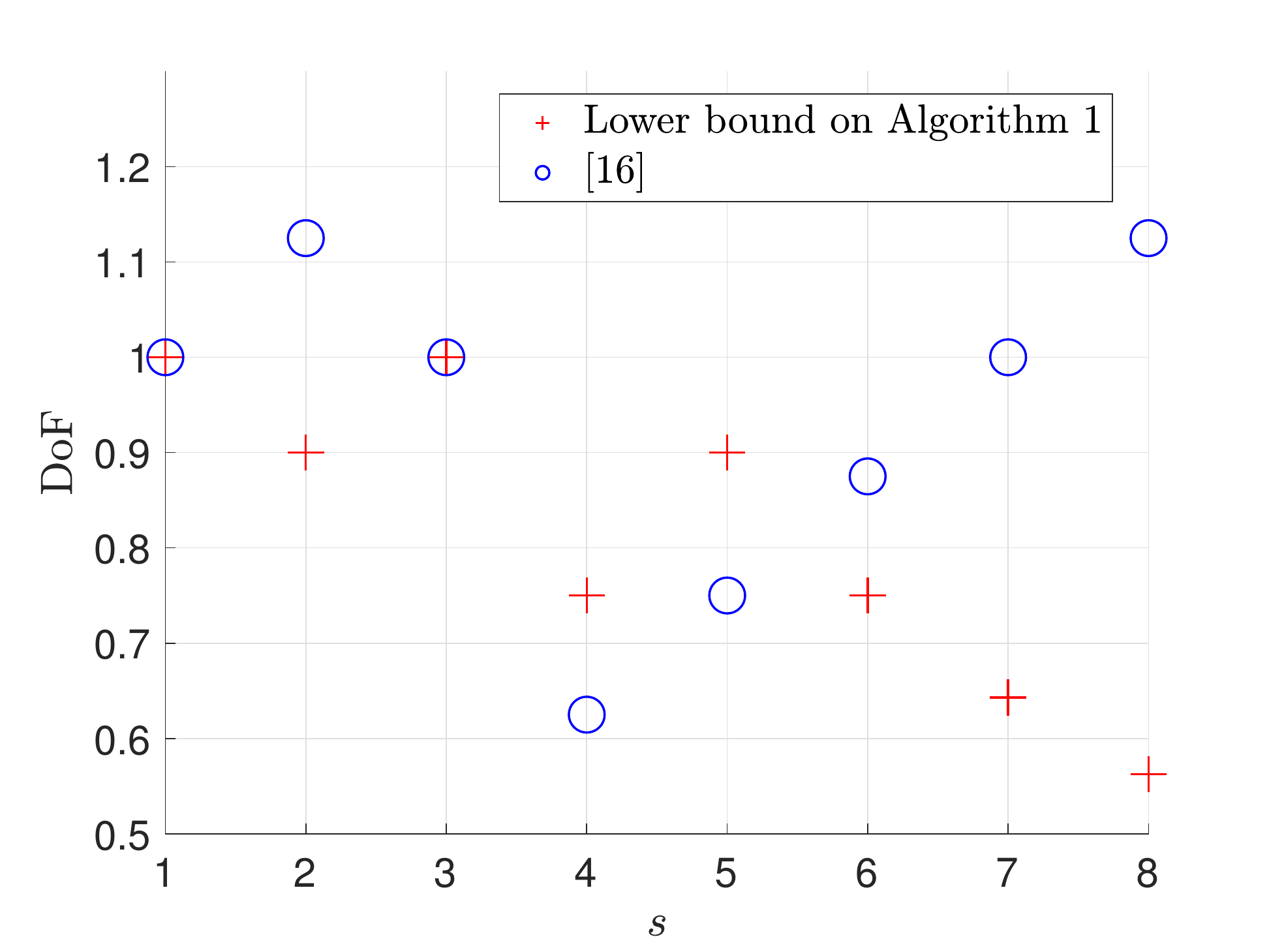}
\end{minipage}%
}%
\subfigure[$N=K=9,M=2.$]{
\begin{minipage}[t]{0.33\linewidth}
\centering
\includegraphics[scale=0.3]{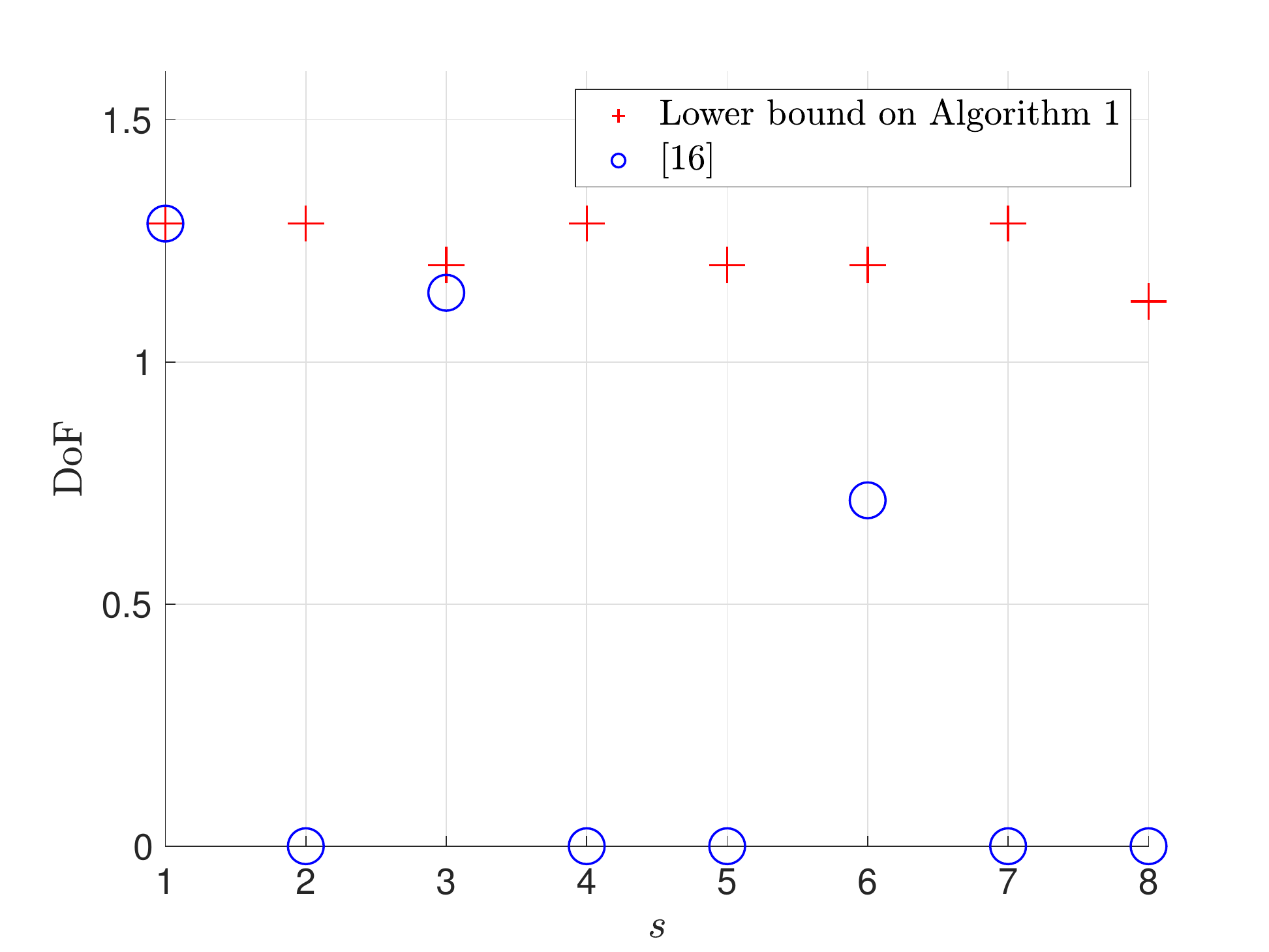}
\end{minipage}%
}%
\centering
\subfigure[$N=K=9,M=3.$]{
\begin{minipage}[t]{0.33\linewidth}
\centering
\includegraphics[scale=0.3]{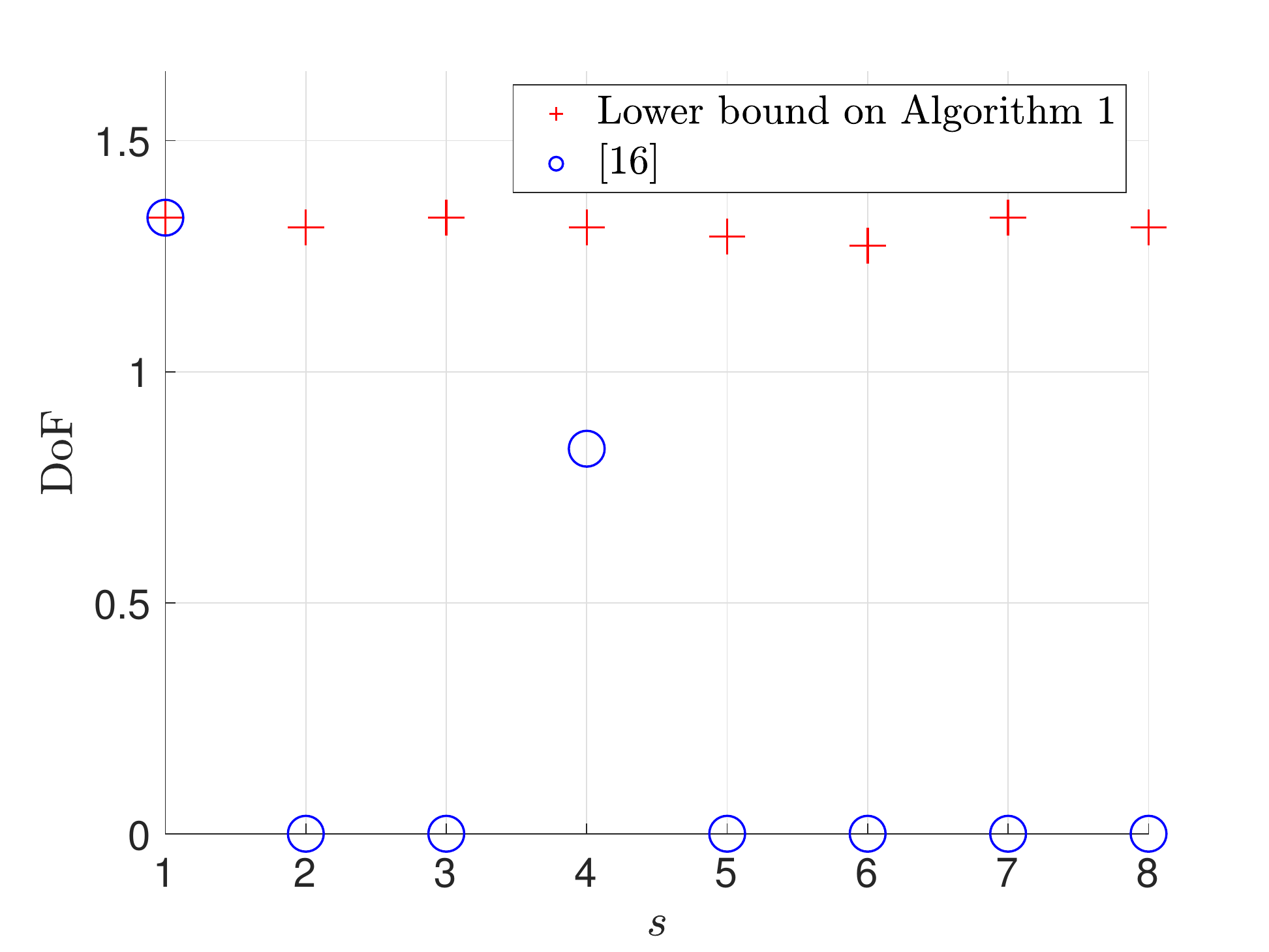}
\end{minipage}%
}

\caption{DoF lower bound for the proposed greedy scheme in Algorithm~\ref{alg:proposed} compared to the DoF of the scheme in \cite{8950279} as a function of $s$ for various network parameters.}
\label{fig:DoF}
\end{figure*}
Moreover, it is guaranteed in Algorithm~\ref{alg:proposed} that each of the $K$ users decodes at least $s-1$ messages in each time slot, except for the final time slot. Therefore, the number of time slots is also bounded by $\left\lceil\frac{\binom{K-1}{t}}{s-1}\right\rceil+1$.
Consequently, we have
\begin{align}\label{UpperBoundBAlg}
B_u \triangleq \text{min} \left\{
\left\lceil \frac{\binom{K}{t+1}}{s\left\lfloor {\frac{K}{{t + 1}}} \right\rfloor} \right\rceil ,
\left\lceil\frac{\binom{K-1}{t}}{s-1}\right\rceil+1
\right\},
\end{align}
and the DoF of the proposed low-complexity scheme satisfies
\begin{align}\label{DoFLowerBoundBu}
\mathrm{DoF} \ge \frac{\binom{K}{t}}{sB_u}.    
\end{align}

The lower bound on the DoF of the proposed greedy scheme is depicted in Fig.~\ref{fig:DoF} for different network parameters, in comparison with the scheme in \cite{8950279}.
As the greedy scheme in Algorithm~\ref{alg:proposed}, a low-complexity scheme that achieves the DoF lower bound can always be obtained for any $s$, while the scheme in \cite{8950279} may not exist for some values of $s$, for which the DoF is set to 0 in Fig.~\ref{fig:DoF}.
Note that the achievable DoF in \cite{8950279} monotonically increases with $\alpha$. For any $s$ value, if an integer $\beta$ is found such that $\binom{t+\beta-1}{t}=s$, then we find the maximum possible $\alpha$ such that $t+\alpha$ is divided by $t+\beta$, yielding the highest DoF of the scheme in \cite{8950279} as depicted in Fig. 3.
It can be seen from Fig.~\ref{fig:DoF} that, the proposed greedy scheme can outperform \cite{8950279} for certain values of $s$, especially for small $s$ that is of particular interest in practice.
We  remark that, the derived lower bound on DoF can be loose due to the floor and ceiling operations in (\ref{UpperBoundBAlg}), and the proposed greedy scheme may achieve a DoF strictly higher than the lower bound illustrated here. For instance,
in the scenario of $N=K=9$ and $M=1$ as considered in Fig. 3(d),
the total number of messages for each user to decode is 8. The proposed low-complexity scheme for $s=8$ delivers all the coded messages in only one time slot, achieving the same DoF of $\frac{9}{8}$ as the scheme in \cite{8950279}. Hence,  the lower bound on DoF at $s=8$ is very loose as can be seen in Fig. 3(d).
We finally remark that, while the DoF lower bound is in general loose, it is tight at certain values of $s$ regardless of the values of $N, K$, and $M$. Specifically, when $s=1$, or when $s$ is sufficiently large such that $B_u=1$, the scheme obtained via the relaxed algorithm, which derives the DoF lower bound in (\ref{DoFLowerBoundBu}), is identical to the one obtained via Algorithm~\ref{alg:proposed} with equal blocklength allocation over all the time slots; and hence, the right-hand-side term in (\ref{DoFLowerBoundBu}) is the exact DoF achieved by the scheme in Algorithm~\ref{alg:proposed} when simply allocating equal blocklength across all the time slots.

\section{Joint Optimization of Beamforming and Coded Content Delivery}\label{sparsity}

In this section, we formulate a sparsity constrained power minimization problem to jointly optimize the beamformers and the content delivery scheme. The sparsity induced problem directly limits the number of messages to be decoded by each user at any time slot, and the indicator function $v_\mathcal{T}(i)$ is identified by setting $v_\mathcal{T}(i) = |R^\mathcal{T}(i)|_0,~\text{for}~\forall i,\mathcal{T}$, where $|\cdot|_0$ denotes the $\ell_0$-norm and is equal to the number of non-zero elements of a vector. Therefore, we impose an $\ell_0$-norm constraint on the rates of messages at any time slot $i$ as follows:
\begin{align}
\sum\limits_{\mathcal{T} \in \mathcal{S}_k} \vert R^\mathcal{T}(i) \vert_0 \leq s,~\forall k,i.
\end{align}
In this section, we assume equal blocklength allocation over all the $B$ time slots for simplicity.
Then, the minimum required power problem with the constrains on the number of messages to be decoded by any user at any time slot can be formulated as follows:
\begin{subequations}\label{eq:optimization_sparse}
\begin{flalign}
\min\limits_{\{\bm{w}_\mathcal{T}(i)\},\{R^\mathcal{T}(i)\}}&~\frac{1}{B}\sum\limits_{i=1}^B \ \ \!\!\!\! \sum\limits_{\mathcal{T} \in \mathcal{S}} \ \Vert\bm{w}_{\mathcal{T}}(i)\Vert^2\\
\text{s.t.} & ~~\sum\limits_{\mathcal{T} \in \pi_{\mathcal{S}_k}^j} R^\mathcal{T}(i) \leq \frac{1}{B}~\!\text{log}_2 \bigg(1+\sum\limits_{\mathcal{T} \in \pi_{\mathcal{S}_k}^j}\gamma_k^\mathcal{T}(i) \bigg),\nonumber\\
&~~~~~~~~~~~~~~~~~~~~~~\;\forall \pi_{\mathcal{S}_k}^j \in \Pi_{\mathcal{S}_k},~\forall k,i,\label{eq:optimization_sparse_c1}\\
& ~~\sum\nolimits_{i=1}^B R^\mathcal{T} (i) \geq \frac{R}{\binom{K}{t}},~\forall \mathcal{T},\label{eq:optimization_sparse_c3}\\
&\sum\limits_{\mathcal{T} \in \mathcal{S}_k} \vert R^\mathcal{T}(i) \vert_0 \leq s,~\forall k,i,\label{eq:optimization_sparse_c2}
\end{flalign}
\end{subequations}
where $\gamma_k^\mathcal{T}(i)$ is defined in (\ref{eq:SINR}). In this problem, the objective is to minimize the average transmission power over all the time slots; constraints (\ref{eq:optimization_sparse_c1})-(\ref{eq:optimization_sparse_c3}) guarantee successful decoding of all the required messages at each user in each time slot; and constraint (\ref{eq:optimization_sparse_c2}) limits the number of messages decoded by each user in any time slot.
Since (\ref{eq:optimization_sparse_c2}) limits only the number of messages decoded by each user in any time slot, without assuming any specific content delivery scheme, the problem in (\ref{eq:optimization_sparse}) includes the content delivery schemes in \cite{8950279} as a special case.
This formulation also generalizes the one presented in Section III when the time slots are of equal duration. However, note that the number of time slots $B$ is a free variable for the greedy scheme, while it is assumed to be given in (\ref{eq:optimization_sparse}).

To deal with the discontinuous $\ell_0$-norm constraint in (\ref{eq:optimization_sparse_c2}), we approximate it with a differentiable continuous function \cite{Rinaldi2010}
\begin{align}\label{eq:appoximation}
f(R^\mathcal{T}(i),t) \triangleq \frac{2}{\pi}\text{arctan}\frac{R^\mathcal{T}(i)}{\xi},
\end{align}
where $\xi > 0$ is a prescribed constant that determines the approximation accuracy. The function in (\ref{eq:appoximation}) is concave w.r.t. $R^\mathcal{T}(i)$, therefore the approximate constraint for (\ref{eq:optimization_sparse_c2})
\begin{align}\label{eq:constrant_approximation}
\sum\limits_{\mathcal{T} \in \mathcal{S}_k} f(R^\mathcal{T}(i),t) \leq s,~\forall k,i,
\end{align}
is concave and can be treated as a difference of convex function.
Overall, the original problem in (\ref{eq:optimization_sparse}) can be approximated by the following problem:
\begin{subequations}\label{eq:optimization_overall_p3}
\begin{flalign}
\min\limits_{\{\bm{w}_\mathcal{T}(i)\},\{R^\mathcal{T}(i)\}}&~\sum\limits_{i=1}^B \ \ \!\!\!\! \sum\limits_{\mathcal{T} \in \mathcal{S}} \ \frac{1}{B} \Vert\bm{w}_{\mathcal{T}}(i)\Vert^2\\
\text{s.t.}&(\text{\ref{eq:optimization_sparse_c1}}),(\text{\text{\ref{eq:optimization_sparse_c3}}})\text{ and}~(\text{\ref{eq:constrant_approximation}}).\nonumber
\end{flalign}
\end{subequations}
Similarly to (\ref{eq:power}), the problem in (\ref{eq:optimization_sparse}) can be solved via the SCA method.
Specifically, we introduce $\eta_{\pi_{\mathcal{S}_k}}^j(i)\triangleq \sum\limits_{\mathcal{T} \in \pi_{\mathcal{S}_k}^j}\gamma_k^\mathcal{T}(i) $, then the constraint in (\ref{eq:optimization_sparse_c1}) can be rewritten similarly to (\ref{eq:optimization_overall_p2_c4}) and (\ref{eq:optimization_overall_p2_c1}), given by
\begin{align}
&\sum_{\mathcal{T} \in \pi_{\mathcal{S}_k}^j} R^\mathcal{T}(i)\leq \frac{1}{B} \log_2(1+ \eta_{\pi_{\mathcal{S}_k}^j}\!\!\!(i)),~\forall \pi_{\mathcal{S}_k}^j \in \Pi_{\mathcal{S}_k},\forall k,i,\\
&\sum\nolimits_{\mathcal{I} \in \mathcal{S}_k^C} \vert \bm{h}_k^H \bm{w}_\mathcal{I}(i) \vert^2 - \frac{\sum_{\mathcal{T} \in \pi_{\mathcal{S}_k}^j}\vert \bm{h}_k^H \bm{w}_\mathcal{T}(i) \vert^2}{\eta_{\pi_{\mathcal{S}_k}^j}\!\!\!(i)} \nonumber\\
&~~~~~~~~~~~~~~~~~~~~~~~~+\sigma_k^2\leq 0,~\forall \pi_{\mathcal{S}_k}^j \in \Pi_{\mathcal{S}_k},\forall k,i,
\end{align}
where the constraints in (25) are convex, and the constraints in (26) are in the form of  difference of convex functions. In the $(\nu+1)$-th iteration of the SCA algorithm, the constraints in (26) can be linearized with the first order Taylor expansion at $\{\bm{w}^\nu_\mathcal{T}(i)\}$ and $\{\eta^\nu_{\pi_{\mathcal{S}_k}^j}(i)\}$,
leading to stricter constraints given by
{\footnotesize
\begin{align}
&\sum\nolimits_{\mathcal{I} \in \mathcal{S}_k^C} \vert \bm{h}_k^H \bm{w}_\mathcal{I}(i) \vert^2 + \frac{\sum_{\mathcal{T} \in \pi_{\mathcal{S}_k}^j}\vert \bm{h}_k^H \bm{w}_\mathcal{T}^\nu(i) \vert^2}{\eta^{\nu^2}_{\pi_{\mathcal{S}_k}^j}\!\!\!(i)}\eta_{\pi_{\mathcal{S}_k}^j}\!\!(i)\nonumber\\
&~~~~-\frac{2\sum_{\mathcal{T} \in \pi_{\mathcal{S}_k}^j} \bm{w}_\mathcal{T}^{\nu^H}(i)\bm{h}_k\bm{h}_k^H \bm{w}_\mathcal{T}(i)}{\eta^{\nu}_{\pi_{\mathcal{S}_k}^j}\!\!\!(i)}
+\sigma_k^2\leq 0,~\forall \pi_{\mathcal{S}_k}^j \in \Pi_{\mathcal{S}_k},\forall k,i,
\end{align}}
where $\{\bm{w}^\nu_\mathcal{T}(i)\}$ and $\{\eta^\nu_{\pi_{\mathcal{S}_k}^j}(i)\}$ are the solutions to the subproblem in the $\nu$-th iteration.
The same linearization technique can also be performed for the constraints in (23), yielding stricter constrains given by
\begin{align}
    &\quad \sum\limits_{\mathcal{T} \in \mathcal{S}_k}\text{arctan}\left( \frac{R^{\mathcal{T}^\nu}(i)}{t}\right) + \frac{t}{\left(t^2+R^{\mathcal{T}^{\nu^2}}(i)\right)}\left(R^{\mathcal{T}}(i) - R^{\mathcal{T}^\nu}(i)\right) \nonumber\\
    &~~~\leq \frac{\pi s}{2},~\forall k,i,
\end{align}
where $\{R^{\mathcal{T}^\nu}(i)\}$ are the solutions to the subproblem in the $\nu$-th iteration.
Overall, the convex subproblem to be solved in the $(\nu+1)$-th iteration is
\begin{subequations}\label{eq:optimization_sca_sparse}
\begin{flalign}
&\min\limits_{\{\bm{w}_\mathcal{T}(i)\},\{R^\mathcal{T}(i),\{\eta_{\pi_{\mathcal{S}_k}^j}\!\!\!(i)\}}~\frac{1}{B}\sum\limits_{i=1}^B \ \ \!\!\!\!  \sum\limits_{\mathcal{T} \in \mathcal{S}} \ \Vert\bm{w}_{\mathcal{T}}(i)\Vert^2\\
&~~~~~~~~~~\text{s.t.}
~~~~~~~~~~~\text{(21c}),~\text{(25), (27), \text{and}~\!(28)}\nonumber.
\end{flalign}
\end{subequations}
The initialization of the SCA algorithm for problem (\ref{eq:optimization_sca_sparse}) for a given value of $s$ requires a content delivery scheme with less or equal complexity, which can be obtained via Algorithm~\ref{alg:proposed} in Section IV. The associated beamforming design can be readily obtained similarly to (\ref{eq:optimization_sca_p3}).
From problem (\ref{eq:optimization_sparse}), we can also conclude that the minimum required power of a content delivery scheme is a non-decreasing function of $s$, as the problem becomes more relaxed as $s$ increases.

\section{Simulation Results}
We consider a single-cell with radius 500m, and users uniformly randomly distributed in the cell. Channel vectors $\bm{h}_{k}$ are written as $\bm{h}_k=(10^{{-\text{PL}}/10})^{1\slash 2}\bm{\tilde{h}}_k$, $\forall k$, where $\bm{\tilde{h}}_k$ denotes an i.i.d. vector accounting for Rayleigh fading of unit power, and the path loss exponent is modeled as $\text{PL}=148.1+37.6\text{log}_{10}(v_k)$, with $v_k$ denoting the distance between the BS and the user (in kilometers). The noise variance is set to $\sigma^2_k=\sigma^2=-134$ dBW for all the users.
Throughout this section, we assume that the number of transmit antennas is $N_T\geq K-t$  as required for achieving a satisfactory DoF performance.
All simulation results are averaged over 300 independent trials computed with CVX \cite{cvx}.

The scheme with $B=1$ time slot will be referred to as the full superposition (FS) scheme. FS has the best performance in terms of transmit power given enough spatial DoF, and serves as a baseline, but it also has the highest complexity.
To compare our results with those in \cite{8950279}, same number of coded messages are transmitted to each user in each time slot for both schemes. We note here that with the use of $\beta$ parameter, the scheme in \cite{8950279} can be improved by serving disjoint subsets of users simultaneously without increasing the complexity, but the improvement is only applicable when the size of user subset can be partitioned equally and exactly. Therefore, the scheme in \cite{8950279} cannot handle certain settings such as the case of $s=2$ in Example 1.

\begin{figure}[!tp]
\centering
\includegraphics[scale = 0.5]{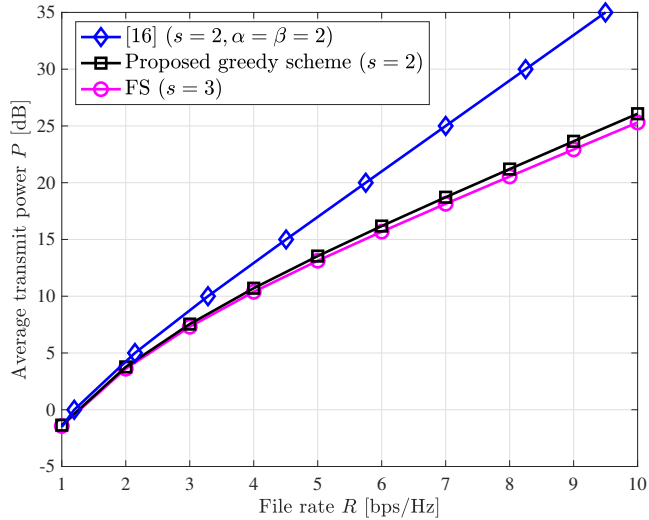}
\caption{Average transmit power $P$ as a function of rate $R$ for $N=K=5$, $M=1$, and $N_T=6$.}
\label{PvR1}
\end{figure}

We first present the average transmit power as a function of the target rate $R$ in Fig.~\ref{PvR1} for Example~1, assuming that the BS is equipped with $N_T=6$ antennas.
The scheme in \cite{8950279} that satisfies $s=2$ is adopted for fair comparison, where $t+\alpha=3$ users are served in each time slot.
We observe that the proposed greedy scheme provides significant savings in the transmit power compared to \cite{8950279} at all rates.
The power savings increase with rate $R$ as a result of the increased superposition coding gain. Furthermore, the gap between the proposed greedy scheme and FS is quite small, and remains almost constant with rate. At $R=8~ \text{bps/Hz}$, the power loss of the scheme in \cite{8950279} and ours compared to FS are about 8.5 dB and 0.5 dB, respectively.
Hence, we can conclude that the proposed greedy scheme provides significant reduction in the computational complexity without sacrificing the performance much.

\begin{figure}[!tp]
\centering
\includegraphics[scale = 0.5]{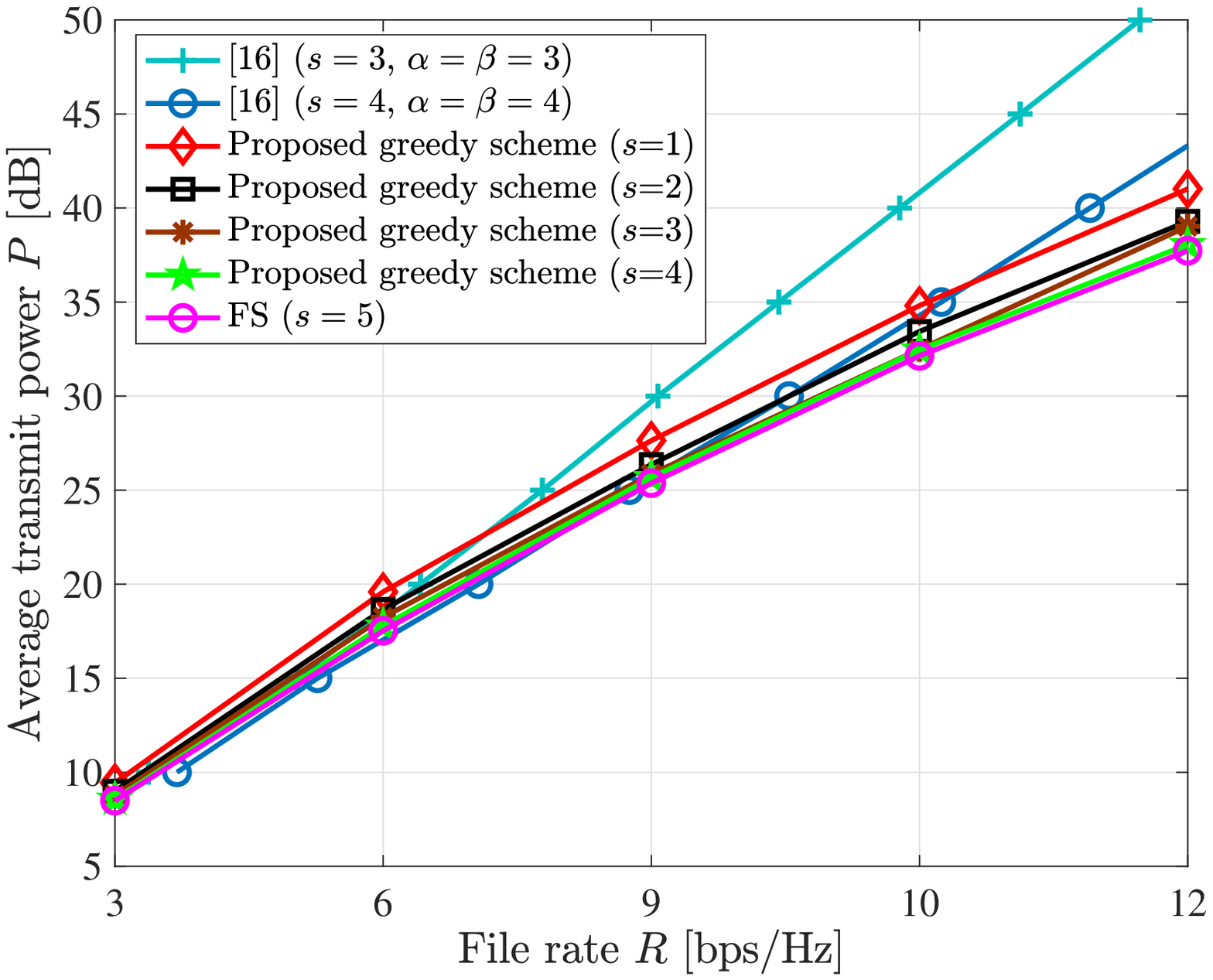}
\caption{Average transmit power $P$ as a function of rate $R$ for $N=K=6$, $M=1$, and $N_T=6$.}
\label{fig:power_n6}
\end{figure}

The average transmit power as a function of file rate $R$ is further investigated for the setting with $N=6$ files, $K=6$ users, $M=1$, and $N_T=6$ antennas. Similarly to Fig.~\ref{PvR1}, it is observed in Fig.~\ref{fig:power_n6} that our proposed low-complexity greedy scheme substantially outperforms the scheme in \cite{8950279} with the same value of $s$ in the high SNR regime. For example, the power savings of the greedy delivery scheme compared to  \cite{8950279} are 8dB and 2dB, for $s=3$ and $s=4$, respectively, at $R=10$ bps/Hz. The power gain is again observed to be larger as the rate increases, while in the low SNR/rate regime, all the schemes achieve comparable performance regardless of $s$.
Also, for the proposed greedy scheme, a larger $s$ allows achieving the same rate with lower transmit power in the high SNR regime, at the expense of increased complexity at the receivers. It is noted that the proposed greedy scheme yields the same content delivery scheme in terms of the transmitted coded messages in each time slot as the one in \cite{8950279} when $s=1$ and $s=2$, which correspond to $\alpha=5,\beta=1$, and $\alpha=5,\beta=2$ in \cite{8950279}, respectively.

\begin{figure}[!tp]
\centering
\includegraphics[scale = 0.5]{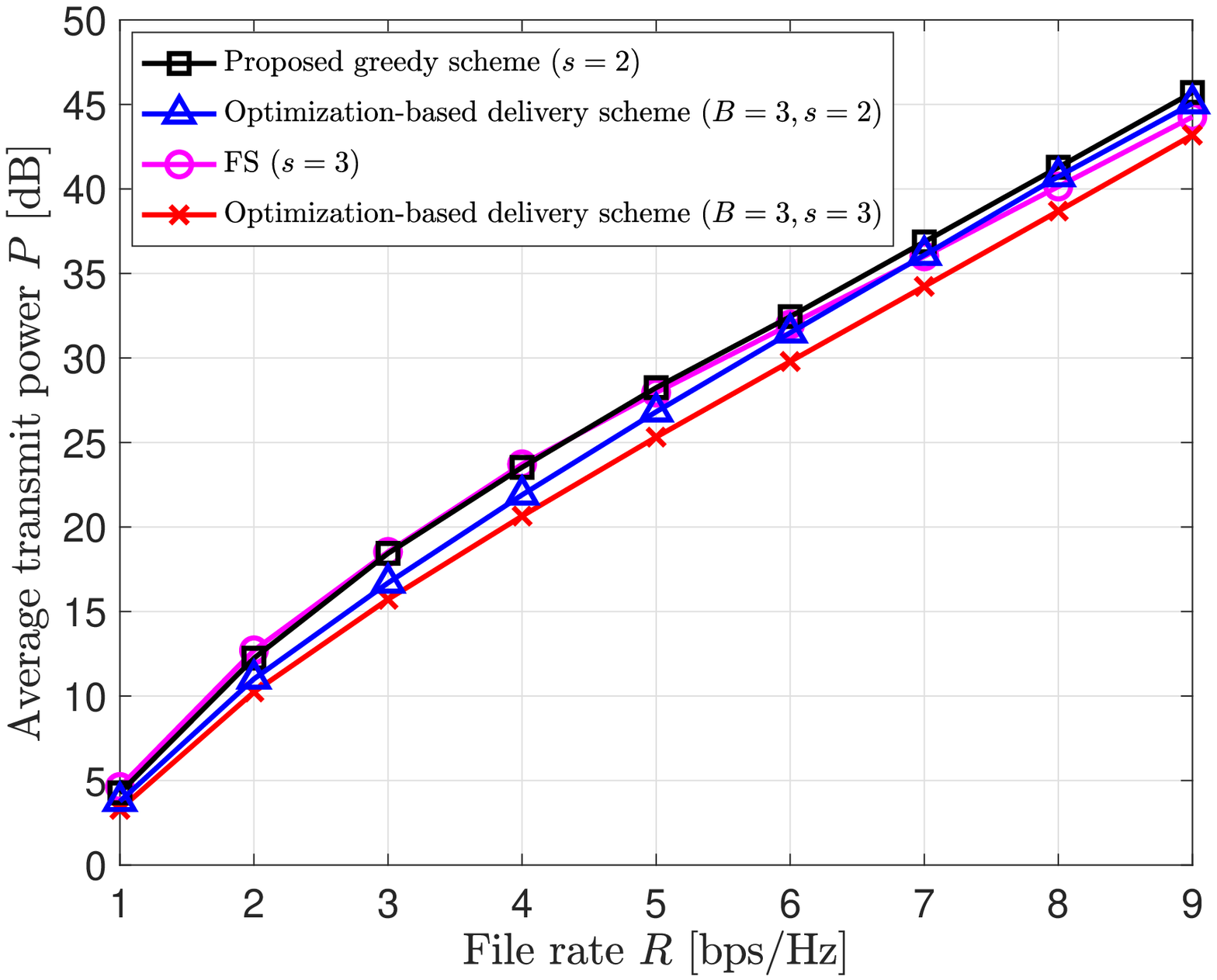}
\caption{Average transmit power $P$ as a function of rate $R$ for $N=K=4$, $M=1$, and $N_T=3$.}
\label{PvRn4}
\end{figure}

\begin{figure}[!tp]
\centering
\includegraphics[scale = 0.5]{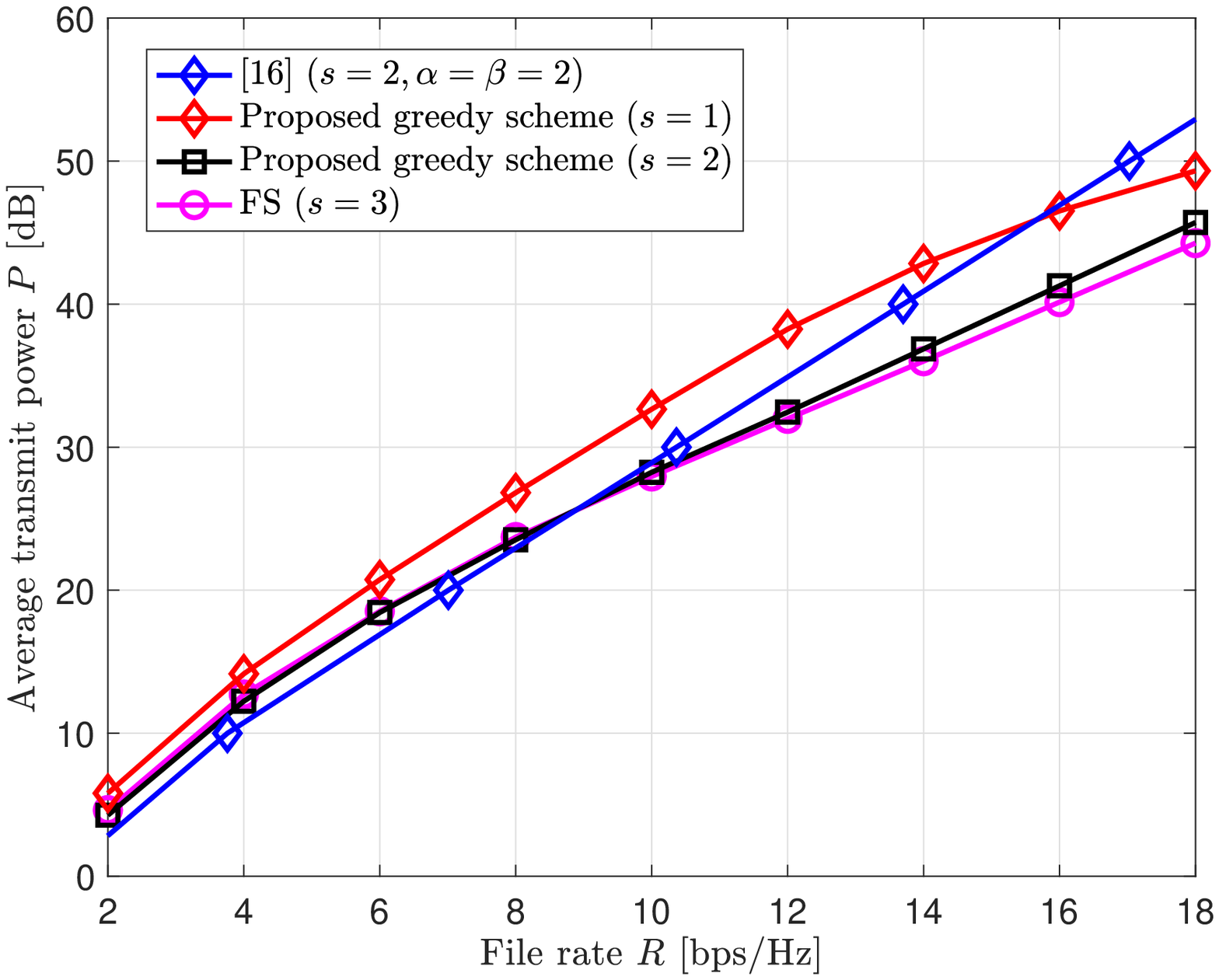}
\caption{Average transmit power $P$ as a function of rate $R$ for $N=K=4$, $M=1$, and $N_T=3$.}
\label{PvR2}
\end{figure}

Fig.~\ref{PvRn4} and Fig.~\ref{PvR2} show the average transmit power versus rate $R$ for Example 2, with $N=4$ files, $K=4$ users, $M=1$, and $N_T=3$ antennas.
In Fig.~\ref{PvRn4}, we compare our greedy content delivery scheme with the one obtained by solving the problem in (\ref{eq:optimization_sparse}) for $s=2$ and $s=3$, by setting $B=3$. It is seen that the greedy scheme can achieve comparable performance, and the performance gap is small especially for high rates. The optimization-based content delivery scheme with $s=3$ is found to outperform the one with $s=2$ as expected, and the improvement is larger as the rate increases.
In Fig.~\ref{PvR2}, it is interesting to see that when the rate is low, the scheme in \cite{8950279} slightly outperforms both the FS and the proposed schemes.
A similar observation has been made in \cite{8950279} showing that a higher rate can be achieved when transmitting a smaller number of coded messages at low SNR, reducing the interference from coded packets that are not decoded at each user.
Due to insufficient spatial degrees of freedom, both the FS and the proposed schemes fail to manage the interference between data streams.
We conclude that this effect occurs only for low rates, as the benefit of superposition coding becomes more dominant at higher rates.
We note that the greedy scheme coincides with the one in \cite{8950279} for $s=1$ in terms of the transmitted coded messages in each time slot, but this does not always happen. For instance, when $s=2$, the only option in \cite{8950279} to keep the same level of complexity is to serve 3 users in each time slot.

\begin{figure}[!tp]
\centering
\includegraphics[scale = 0.5]{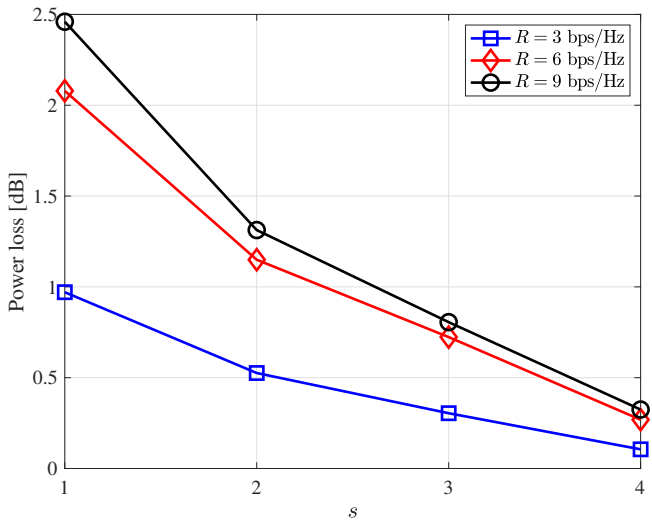}
\caption{Power loss w.r.t. FS as a function of $s$ for $N=K=6$, $M=1$, and $N_T=6$.}
\label{fig:power_loss}
\end{figure}

We plot in Fig.~\ref{fig:power_loss} the power loss of the proposed scheme in Algorithm~\ref{alg:proposed} compared to FS as a function of $s$, that is, how much more the required transmit power is compared to the transmit power required by FS. This results in a more clear plot.
Assuming $N=6$, $K=6$, $M=1$, we let $s$ take values from $\{1,2,3,4,5\}$, where $s=5$ corresponds to the FS scheme in which all the $\binom{K}{t+1}=15$ coded messages are transmitted simultaneously.
FS requires less transmit power at the expense of a high decoding complexity.
When $s=1$, the model boils down to the single-cell multigroup multicasting problem, which has the lowest computation and implementation complexity. In general, Fig.~\ref{fig:power_loss} can be considered as the trade-off curve between the performance and complexity for each rate value, both of them increasing with $s$.

\section{Conclusions} \label{sec_ch4:conclusion}

In this paper, we have studied cache-aided content delivery from a multi-antenna BS in the finite SNR regime.
We have formulated a general beamforming scheme that multicasts coded files over multiple orthogonal time slots. We have then
specialized this general formulation to a low-complexity greedy scheme
by limiting the number of coded messages targeted at each user at each time slot. This scheme provides the flexibility to adjust the computational complexity of the optimization problem and the receiver complexity.
We have then formulated the constraint on the number of coded messages targeted at each user at each time slot as a sparsity constraint, and solved the resulting mixed-integer non-convex optimization problem using the SCA method.
Compared with FS, where all the coded messages are transmitted simultaneously, and the scheme obtained via the sparsity-constrained optimization framework,
the greedy scheme achieves comparable performance, and outperforms the one proposed in \cite{8950279} for all values of SNR and rate with sufficient spatial degrees of freedom, while the improvement is limited to high data rate values when the BS does not have sufficiently many transmit antennas. Furthermore, the gap between the greedy delivery scheme and the optimization-based delivery scheme decreases as the SNR/power increases.
When considering practical implementations, one must choose a suitable value of $s$ that yields an acceptable performance while keeping the complexity feasible.
The satisfactory DoF performance of the proposed low-complexity scheme is guaranteed with at least $K-t$ antennas, while the analysis of overloaded systems with $K$ users served by less than $K-t$ antennas is
left as future work.

\appendices
\section{An  upper bound on the solution of Problem (\ref{eq:power})}
It is noted that the constrains are in the form of  difference of convex functions, which can be approximated by linearizing the concave functions, resulting in a convex problem that can be solved via SCA techniques.
To see this, we first rewrite the problem in (\ref{eq:power}) as
\begin{subequations}\label{eq:optimization_overall_p2}
\begin{flalign}
&\!\!\!\!\min\limits_{\{\bm{w}_\mathcal{T}(i)\},\{R^\mathcal{T}(i),\{\eta_{\pi_{\mathcal{S}_k}^j}\!\!\!(i)\}}~\sum\limits_{i=1}^B \ \ \!\!\!\! \sum\limits_{\mathcal{T} \in \mathcal{S}} \frac{n_i}{n} \Vert\bm{w}_{\mathcal{T}}(i)\Vert^2\\
&~~~~~~~~~~\text{s.t.}
\sum_{\mathcal{T} \in \pi_{\mathcal{S}_k}^j} R^\mathcal{T}(i)\leq \frac{n_i}{n} \log_2(1+ \eta_{\pi_{\mathcal{S}_k}^j}\!\!\!(i)),\nonumber\\
&~~~~~~~~~~~~~~~~~~~~~~~~~~~~~~~~~~\forall \pi_{\mathcal{S}_k}^j \in \Pi_{\mathcal{S}_k},\forall k,i,\label{eq:optimization_overall_p2_c4}\\
&~~~~~~~~~~\sum\nolimits_{\mathcal{I} \in \mathcal{S}_k^C} \vert \bm{h}_k^H \bm{w}_\mathcal{I}(i) \vert^2 - \frac{\sum_{\mathcal{T} \in \pi_{\mathcal{S}_k}^j}\vert \bm{h}_k^H \bm{w}_\mathcal{T}(i) \vert^2}{\eta_{\pi_{\mathcal{S}_k}^j}\!\!\!(i)} \nonumber\\
&~~~~~~~~~~~~~~~~~~~~~~~~~~~+\sigma_k^2\leq 0,~\forall \pi_{\mathcal{S}_k}^j \in \Pi_{\mathcal{S}_k},\forall k,i,\label{eq:optimization_overall_p2_c1}\\
& ~~~~~~~~~~\text{(\ref{eq:constraint2}), \text{(\ref{eq:scheme_constraint3})}}~\text{and}~\text{(\ref{eq:scheme_constraint})},\nonumber
\end{flalign}
\end{subequations}
where $\eta_{\pi_{\mathcal{S}_k}^j}\!\!\!(i) \triangleq \sum\limits_{\mathcal{T} \in \pi_{\mathcal{S}_k}^j}\gamma_k^\mathcal{T}(i) $. The constraint in (\ref{eq:optimization_overall_p2_c1}) is  the difference of convex function, since $\sum_{\mathcal{T} \in \pi_{\mathcal{S}_k}^j}\vert \bm{h}_k^H \bm{w}_\mathcal{T}(i) \vert^2\slash\eta_{\pi_{\mathcal{S}_k}^j}\!\!\!(i)$ is the sum of quadratic-over-linear functions of $\bm{w}_\mathcal{T}(i)$ and $\eta_{\pi_{\mathcal{S}_k}^j}\!\!\!(i)$.
Therefore, a sequence of convex subproblems can be solved iteratively to approximately tackle this convex-concave problem \cite{6788812}, with the subproblem in the $(\nu+1)$-th iteration given by
\begin{subequations}\label{eq:optimization_sca_p3}
\begin{flalign}
&\min\limits_{\{\bm{w}_\mathcal{T}(i)\},\{R^\mathcal{T}(i),\{\eta_{\pi_{\mathcal{S}_k}^j}\!\!\!(i)\}}~\sum\limits_{i=1}^B \ \ \!\!\!\!  \sum\limits_{\mathcal{T} \in \mathcal{S}} \frac{n_i}{n} \Vert\bm{w}_{\mathcal{T}}(i)\Vert^2\\
&\text{s.t.}~ \sum_{\mathcal{T} \in \pi_{\mathcal{S}_k}^j} R^\mathcal{T}(i)\leq \frac{n_i}{n} \log_2(1+ \eta_{\pi_{\mathcal{S}_k}^j}\!\!\!(i)),~\forall \pi_{\mathcal{S}_k}^j \in \Pi_{\mathcal{S}_k},\forall k,i,\label{eq:optimization_sca_p3_c4}\\
&~~~~\sum\nolimits_{\mathcal{I} \in \mathcal{S}_k^C} \vert \bm{h}_k^H \bm{w}_\mathcal{I}(i) \vert^2 + \frac{\sum_{\mathcal{T} \in \pi_{\mathcal{S}_k}^j}\vert \bm{h}_k^H \bm{w}_\mathcal{T}^\nu(i) \vert^2}{\eta^{\nu^2}_{\pi_{\mathcal{S}_k}^j}\!\!\!(i)}\eta_{\pi_{\mathcal{S}_k}^j}\!\!(i)\nonumber\\
&~~-\frac{2\sum_{\mathcal{T} \in \pi_{\mathcal{S}_k}^j} \bm{w}_\mathcal{T}^{\nu^H}(i)\bm{h}_k\bm{h}_k^H \bm{w}_\mathcal{T}(i)}{\eta^{\nu}_{\pi_{\mathcal{S}_k}^j}\!\!\!(i)}
+\sigma_k^2\leq 0,~\forall \pi_{\mathcal{S}_k}^j \!\!\in\! \Pi_{\mathcal{S}_k},\!\forall k,i,\label{eq:optimization_sca_p3_c1}\\
&~~~~~\text{(\ref{eq:constraint2}),~\text{(\ref{eq:scheme_constraint3})}}~\text{and}~ \text{(\ref{eq:scheme_constraint})},\nonumber
\end{flalign}
\end{subequations}
given the solution of $\bm{w}^\nu_\mathcal{T}(i)$, $R^{\mathcal{T}^\nu}(i)$, and $\eta^\nu_{\pi_{\mathcal{S}_k}^j}(i)$ obtained in the $\nu$-th SCA iteration.
Each of the convex subproblems can be efficiently solved with standard interior-point algorithms or off-the-shelf solvers, and the SCA approach is guaranteed to converge to a stationary solution of the original problem in (\ref{eq:power}) \cite{7776948}. Details of the SCA algorithm are outlined in Table.~\ref{tab:SCA Algorithm}.

An initial point in the feasible set of problem (\ref{eq:power}) is required to initialize the SCA algorithm. We first observe that for any feasible target rates $\{R^\mathcal{T}\!(i)~\!\!|\!\!~\forall~\!\!\mathcal{T}~\!\!\!\!\in\!\!\!\!~\mathcal{S}\}_{i=1}^B$ that satisfy the constraints in (\ref{eq:constraint2}) and (\ref{eq:scheme_constraint}), the problem in (\ref{eq:power}) can be decoupled and decomposed into $B$ parallel subproblems, each for a distinct time slot $i \in [B]$, given by
\begin{subequations}\label{eq:optimization_i}
\begin{flalign}
&\{\bm{w}_\mathcal{T}^*(i)\}_{\mathcal{T}\in\mathcal{S}(i)} = \argmin\limits_{\{\bm{w}_\mathcal{T}(i)\}} \ \ \!\!\!\! \sum\limits_{\mathcal{T} \in \mathcal{S}(i)} \ \Vert\bm{w}_{\mathcal{T}}(i)\Vert^2\\
&~~\text{s.t.} \  \ \  \sum\limits_{\mathcal{T} \in \pi_{\mathcal{S}_k}^j} R^\mathcal{T}(i) \leq \frac{n_i}{n}~\!\text{log}_2 \bigg(1+\sum\limits_{\mathcal{T} \in \pi_{\mathcal{S}_k}^j}\gamma_k^\mathcal{T}(i) \bigg),\nonumber\\
&~~~~~~~~~~~~~~~~~~~~~~~~~~~~~~~~~~~~\;\forall \pi_{\mathcal{S}_k}^j \in \Pi_{\mathcal{S}_k},~\forall k,i,\label{eq:constraints}\\
&~~~~~~~\gamma_k^\mathcal{T}(i) = \frac{\vert \bm{h}_k^H \bm{w}_\mathcal{T}(i) \vert^2}{\sum\nolimits_{\mathcal{I} \in \mathcal{S}_k^C} \vert \bm{h}_k^H \bm{w}_\mathcal{I}(i) \vert^2+\sigma_k^2},~\forall k,i,\\
&~~~~~~~\Vert\bm{w}_\mathcal{T}(i)\Vert^2 \geq 0~ \text{for}~\forall v_\mathcal{T}(i) \neq 0,\\
&~~~~~~~\Vert\bm{w}_\mathcal{T}(i)\Vert^2 = 0~ \text{for}~\forall v_\mathcal{T}(i) = 0,
\end{flalign}
\end{subequations}
which is nonconvex. Nevertheless, it can be transformed into a semidefinite programming problem by introducing $\bm{W}_\mathcal{T}(i) \triangleq \bm{w}_\mathcal{T}(i)\bm{w}_{\mathcal{T}}^H(i)$ and dropping the rank-1 constraints on $\bm{W}_\mathcal{T}(i)$, which is given by
\begin{subequations}\label{eq:optimization_SDR}
\begin{flalign}
&\{\bm{W}_\mathcal{T}^*(i)\}_{\mathcal{T}\in\mathcal{S}(i)} = \argmin\limits_{\{\bm{W}_\mathcal{T}(i)\}} \ \ \!\!\!\! \sum\limits_{\mathcal{T} \in \mathcal{S}} \ \text{Tr}\{\bm{W}_{\mathcal{T}}(i)\}\\
&~~\text{s.t.}~\left(2^{\frac{n}{n_i}\sum_{\mathcal{T} \in \pi_{\mathcal{S}_k}^j}\!\!\!\!\!R^\mathcal{T}(i)}-1\right)\left(
\sum\limits_{\mathcal{I} \in \mathcal{S}_k^C} \text{Tr}\{\bm{H}_k \bm{W}_\mathcal{I}(i)\} +\sigma_k^2\right)\nonumber\\
&~~~~~~~~~~~~~~~- \sum_{\mathcal{T} \in \pi_{\mathcal{S}_k}^j}\text{Tr}\{ \bm{H}_k \bm{W}_\mathcal{T}(i)\} \leq 0,~\forall \pi_{\mathcal{S}_k}^j \in \Pi_{\mathcal{S}_k},\forall k,\\
&~~~~~~~\bm{W}_\mathcal{T}(i) \succeq 0, \forall~\!\!v_\mathcal{T}(i)\neq 0,\\
&~~~~~~~\bm{W}_\mathcal{T}(i) = 0, \forall~\!\!v_\mathcal{T}(i)=0,
\end{flalign}
\end{subequations}
and can be efficiently solved with standard interior-point algorithms. However, the solution obtained with semidefinite relaxation is not necessarily rank-1. If the obtained $\bm{W}_\mathcal{T}(i)$'s are all rank-1, then the optimal solution of (\ref{eq:optimization_i}) can be readily recovered from $\bm{W}_\mathcal{T}(i)$. Otherwise, Gaussian randomization can be adopted to obtain a feasible approximation to the optimal solution of (\ref{eq:optimization_i}).
Note that the solution given by (\ref{eq:optimization_i}) is an upper bound on the minimum required power in (\ref{eq:power}) as the rates $\{R^\mathcal{T}\!(i)~\!\!|\!\!~\forall~\!\!\mathcal{T}~\!\!\!\!\in\!\!\!\!~\mathcal{S}\}_{i=1}^B$ are not optimized, which hence can serve as an initial point in the successive convex approximation algorithm to obtain a tighter upper bound on the problem in (\ref{eq:power}).

\begin{table}[!ht]\caption{SCA Algorithm for the Multicast Beamforming Problem with a Given Coded Delivery Scheme}\label{tab:SCA Algorithm}
\centering
\begin{tabular}{l} \hline
STEP 0: Set \(\nu=1\). Set a step size \(\mu\).\\
\quad\quad\quad\ \ \ Initialize \(\bm{w}_\mathcal{T}^\nu(i)\), \(R_\mathcal{T}^\nu(i)\), and \(\eta^\nu_{\pi_{\mathcal{S}_k}^j}\!\!\!(i)\) with feasible values\\
STEP 1: If a stopping criterion is satisfied, then STOP\\
STEP 2: Solve the optimization problem in (\ref{eq:optimization_sca_p3})\\
STEP 3: Update \(\bm{w}_\mathcal{T}^{\nu+1}(i)=\bm{w}_\mathcal{T}^\nu(i) + \mu \left(\bm{w}_\mathcal{T}(i)-\bm{w}_\mathcal{T}^\nu(i)\right)\),\\
\quad\quad\quad\
\ \(R_\mathcal{T}^{\nu+1}(i)=R_\mathcal{T}^\nu(i) + \mu \left(R_\mathcal{T}(i)-R_\mathcal{T}^\nu(i)\right)\),\\
\quad\quad\quad\ \ \(\eta^{\nu+1}_{\pi_{\mathcal{S}_k}^j}\!\!(i)=\eta_{\pi_{\mathcal{S}_k}^j}\!\!\!(i) + \mu \left(\eta_{\pi_{\mathcal{S}_k}^j}\!\!\!(i)-\eta^\nu_{\pi_{\mathcal{S}_k}^j}\!\!\!(i)\right)\),\\
STEP 4: Set \(\nu=\nu+1\), and go to STEP 1\\
\hline
\end{tabular}
\end{table}

\bibliographystyle{IEEEtran}
\bibliography{FINAL VERSION.bbl}
\begin{IEEEbiography}[{\includegraphics[width=1in,height=1.25in,clip,keepaspectratio]{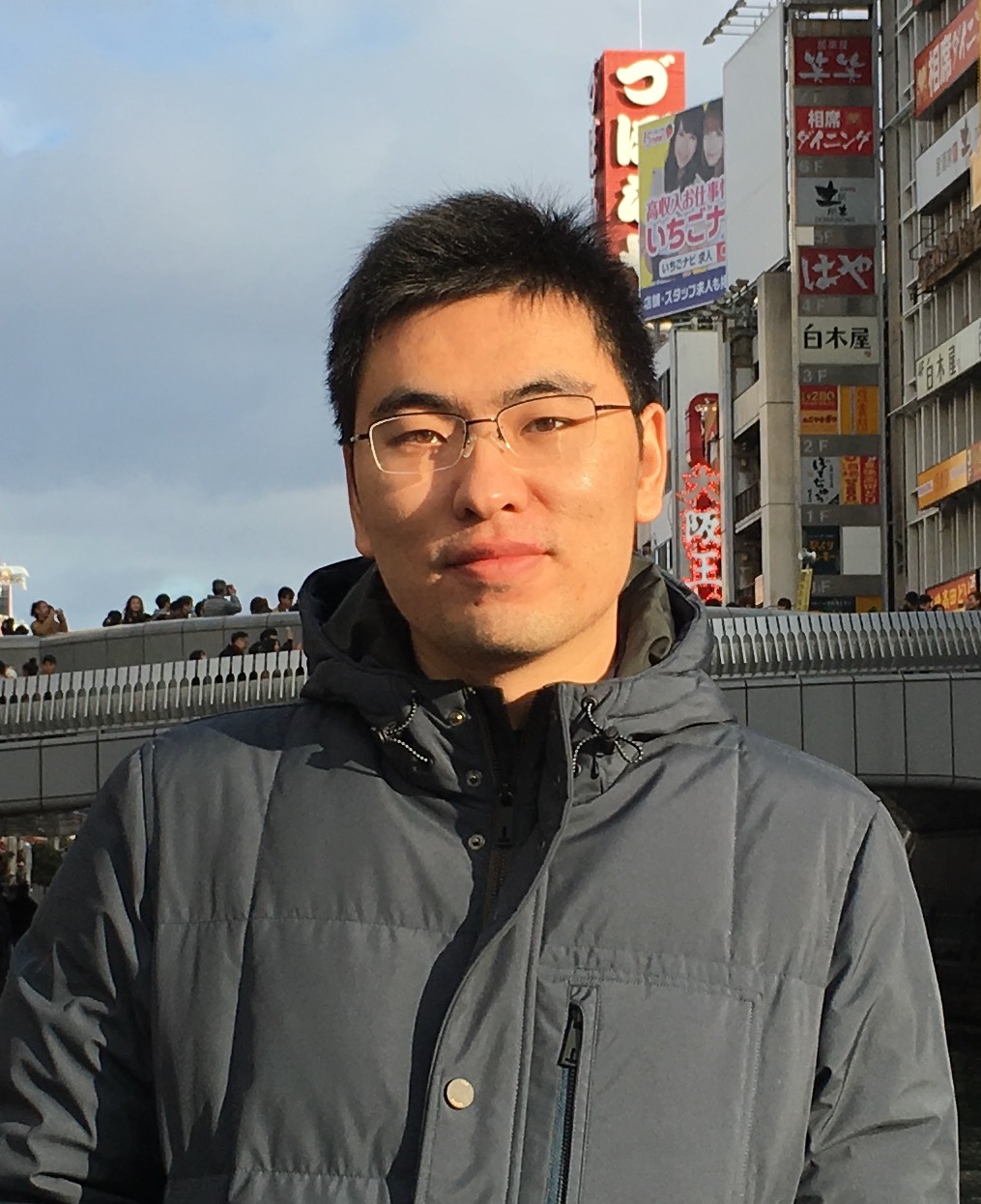}}]{Junlin Zhao}
[S'14-M'19] received the B.Eng. degree in Information Engineering from the Beijing Institute of Technology, Beijing, China, in 2013, and the Ph.D. degree in Electrical Engineering from Imperial College London, London, UK, in 2020. From August 2016 to January 2017, he was a visiting student at the Department of Electrical and Computer Engineering, the Ohio State University, Columbus, Ohio, USA. He is currently a Postdoctoral Researcher at the School of Science and Engineering, the Chinese University of Hong Kong, Shenzhen, P.R. China. His research interests lie in wireless communications and machine learning.
\end{IEEEbiography}

\begin{IEEEbiography}[{\includegraphics[width=1in,height=1.25in,clip,keepaspectratio]{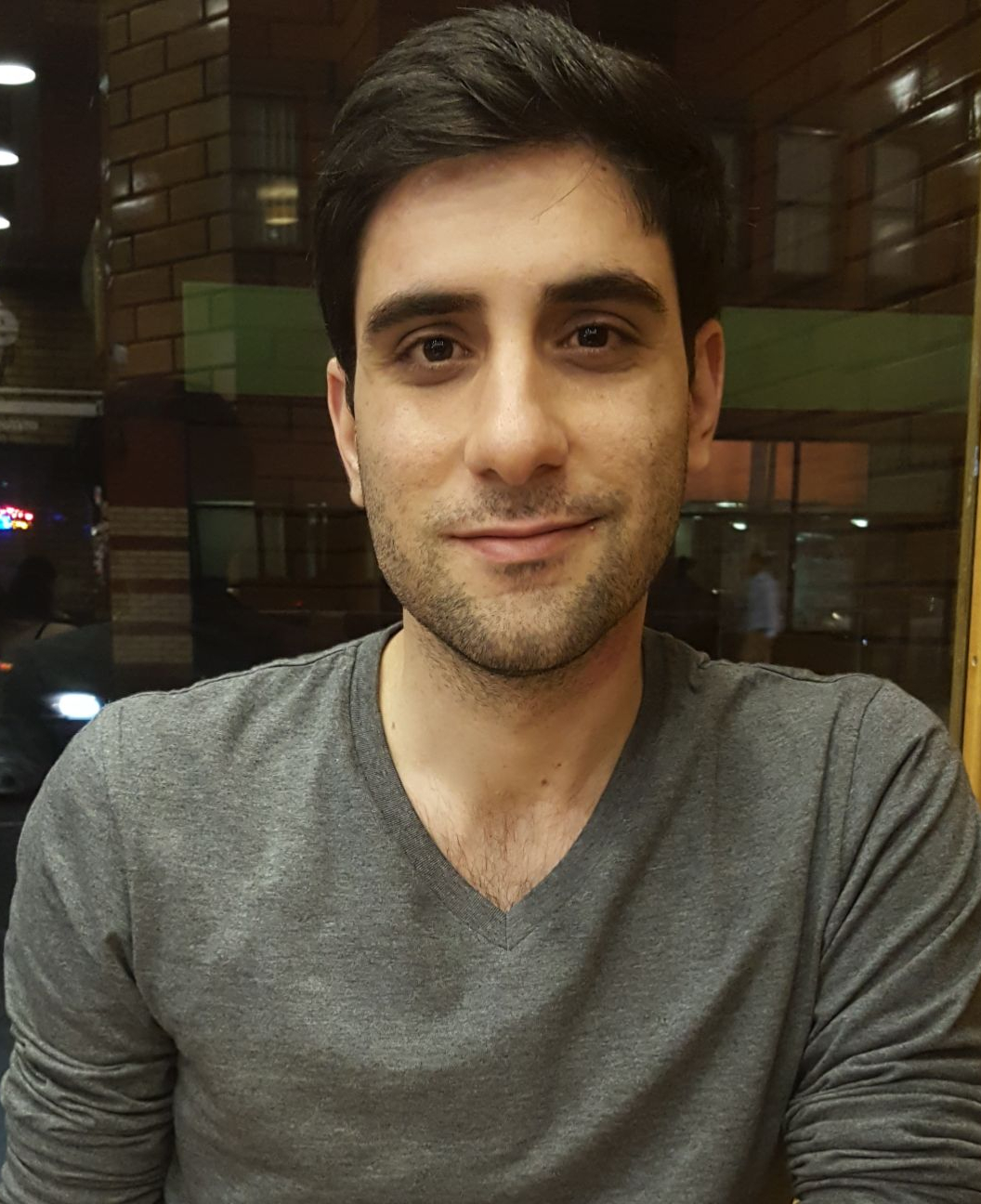}}]{Mohammad Mohammadi Amiri}
[S'16] received his B.Sc. and M.Sc. degrees in Electrical Engineering from Iran University of Science and Technology in 2011 and University of Tehran in 2014, respectively, both with highest rank in classes. He also obtained his Ph.D. degree at Imperial College London in 2019. He is currently a Postdoctoral Research Associate in the Department of Electrical Engineering at Princeton University. His research interests include information and coding theory, machine learning, wireless communications, and signal processing.
\end{IEEEbiography}

\begin{IEEEbiography}[{\includegraphics[width=1in,height=1.25in,clip,keepaspectratio]{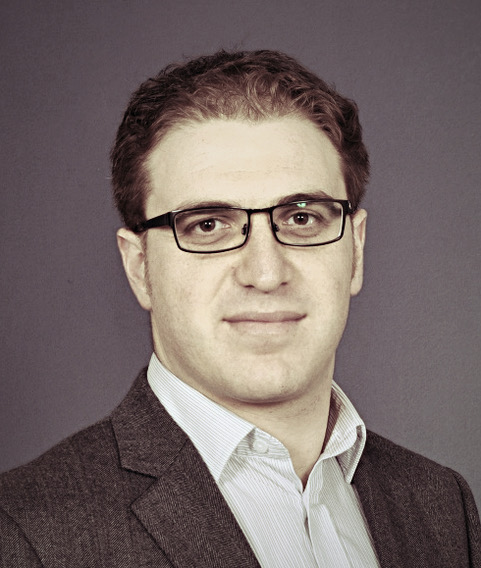}}]{Deniz G\"und\"uz}
[S’03-M’08-SM’13] received the B.S. degree in electrical and electronics engineering from METU, Turkey in 2002, and the M.S. and Ph.D. degrees in electrical engineering from NYU Tandon School of Engineering (formerly Polytechnic University) in 2004 and 2007, respectively. He is currently a Professor in the Electrical and Electronic Engineering Department of Imperial College London, UK. He also serves as the deputy head of the Intelligent Systems and Networks Group, and leads the Information Processing and Communications Laboratory (IPC-Lab). He is also an Associate Researcher at the University of Modena and Reggio Emilia, and held visiting positions at University of Padova (2018, 2019, 2020) and Princeton University (2009-2012). 

His research interests lie in the areas of communications and information theory, machine learning, and privacy. Dr. Gündüz is an Area Editor for the IEEE Transactions on Communications and the IEEE Journal on Selected Areas in Communications (JSAC). He also serves as an Editor of the IEEE Transactions on Wireless Communications. He is a Distinguished Lecturer for the IEEE Information Theory Society (2020-21). He is the recipient of the IEEE Communications Society - Communication Theory Technical Committee (CTTC) Early Achievement Award in 2017, a Starting Grant of the European Research Council (ERC) in 2016, IEEE Communications Society Best Young Researcher Award for the Europe, Middle East, and Africa Region in 2014, Best Paper Award at the 2019 IEEE Global Conference on Signal and Information Processing (GlobalSIP) and  the 2016 IEEE Wireless Communications and Networking Conference (WCNC), and the Best Student Paper Awards at the 2018 IEEE Wireless Communications and Networking Conference (WCNC) and the 2007 IEEE International Symposium on Information Theory (ISIT).
\end{IEEEbiography}

\end{document}